\documentclass[lettersize,journal]{IEEEtran}
\usepackage{amsmath,amsfonts}
\usepackage{algorithmic}
\usepackage{algorithm}
\usepackage{array}
\usepackage[colorlinks,citecolor=blue,linkcolor=blue,urlcolor=blue]{hyperref}
\usepackage[caption=false,font=normalsize,labelfont=rm,textfont=rm]{subfig}
\usepackage{textcomp}
\usepackage{stfloats}
\usepackage{url}
\usepackage{verbatim}
\usepackage{graphicx}
\usepackage{cite}
\usepackage{bm}
\usepackage{cuted}
\usepackage{amssymb}
\usepackage{color}
\usepackage{booktabs}
\usepackage{multirow}
\usepackage{fontspec}
\usepackage{newtxtext}
\usepackage{balance}

\begin{document}

\title{Outage Analysis and Fairness Design for Spatially Correlated FAS-Enabled RSMA Systems}

\author{Jinyuan~Liu,
        Yong~Liang~Guan,~\IEEEmembership{Senior~Member,~IEEE,}
        Tuo~Wu,~\IEEEmembership{Member,~IEEE,}
        Hong~Niu,\\
        Kai-Kit~Wong,~\IEEEmembership{Fellow,~IEEE,}
        and~Bruno~Clerckx,~\IEEEmembership{Fellow,~IEEE}
\thanks{Jinyuan Liu, Hong Niu, and Yong Liang Guan are with the School of Electrical and Electronic Engineering, Nanyang Technological University, Singapore 639798 (e-mail:
jinyuan001@e.ntu.edu.sg; hong.niu@ntu.edu.sg; eylguan@ntu.edu.sg).}
\thanks{Tuo Wu is with the Department of Electrical Engineering City University of Hong Kong, Hong Kong, China. (e-mail: tuo.wu@qmul.ac.uk).}
\thanks{K. K. Wong is affiliated with the Department of Electronic and Electrical Engineering, University College London, Torrington Place, WC1E 7JE, United Kingdom (e-mail: {kai-kit.wong}@ucl.ac.uk) and he is also affiliated with the Department of Electronic Engineering, Kyung Hee University, Yongin-si, Gyeonggi-do 17104, Korea.
}
\thanks{Bruno Clerckx is with the Department of Electrical and Electronic Engineering, Imperial College London, London SW7 2AZ, U.K. (e-mail: b.clerckx@imperial.ac.uk).}
}



\maketitle

\begin{abstract}
Sixth-generation (6G) systems target higher reliability, denser connectivity, and tighter interference control. {Within this context, rate-splitting multiple access (RSMA) is envisioned as a promising candidate to enhance interference management in future wireless networks by flexibly splitting messages into a common and a private part, while fluid antenna systems (FAS) offer the potential to improve spatial selectivity through dynamic port reconfiguration.} Combining RSMA and FAS therefore enables efficient interference control and adaptive antenna utilization in multiuser multi-input single-output (MISO) networks. However, deriving closed-form outage probability (OP) expressions and tractable user fairness optimization in this scenario remains scarce in the literature. This paper studies a multiuser MISO downlink that jointly leverages RSMA and FAS. We develop a spatial correlation model for FAS using block correlation and incorporate linear precoding with zero-forcing and maximum-ratio transmission. Within this model, we derive closed-form OP expressions using a one-factor construction and generalized Gauss-Laguerre quadrature. Building on these expressions, we formulate a fairness objective that minimizes the worst-user OP and propose a low-complexity algorithm with a linear-program feasibility check to obtain the closed-form solution per iteration. Numerical results across different port counts, channel conditions, and target rates validate the analytical analysis, show that FAS-RSMA reduces OP by up to $\mathbf{92\%}$ relative to the fixed-position antenna (FPA) baseline, and demonstrate that fairness-oriented design equalizes user reliability while delivering a $ \mathbf{1}$\,dB SNR gain for the worst user at a fixed outage level. 
\end{abstract}


\section{Introduction}
Sixth-generation (6G) networks aim to support immersive services, ultra-reliable low-latency communications (URLLC), integrated sensing and communications (ISAC), and AI-native applications by delivering higher throughput, reliability, coverage, and connectivity under diverse QoS requirements and dense-interference conditions, where conventional orthogonal multiple-access schemes face inherent limits in spectral efficiency and flexibility \cite{ref1,ref2}.
Rate-splitting multiple access (RSMA) has emerged as a flexible and spectrally efficient multiple-access paradigm for future radio access networks, attracting significant research interest in recent years. It is widely viewed as a promising physical-layer ingredient for 6G systems owing to its ability to accommodate varied service targets \cite{ref3}. The underlying idea of rate-splitting (RS) dates back to \cite{ref4} and was later extended to downlink multi-antenna transmissions in \cite{ref5}. Conceptually, RSMA decomposes each user’s message into a common component and a private component decoded by all users and the intended user, respectively. These decomposed components are mapped to common and private streams and transmitted using linear or nonlinear precoding. At the receivers, successive interference cancellation (SIC) with one or multiple stages separates the common stream from the private ones, implementing the principle of partially decoding interference while treating the remaining multiuser interference as noise \cite{ref6}. Owing to this flexibility, RSMA has been investigated across a wide range of scenarios \cite{ref6,ref7,ref8,ref9,ref10,ref11,ref12,ref13,ref15,ref16,ref17,ref185,ref20}, which consistently demonstrates the RSMA's reliability and flexibility compared with other multiple access schemes. 

Most prior studies on RSMA assume conventional linear (or planar) arrays with fixed inter-element spacing. Such architectures provide spatial diversity, but their scalability is limited by practical constraints \cite{ref20}. In particular, the increasing number of available radio-frequency (RF) chains at user equipment and base stations (BS) aggravates hardware cost, power consumption, and architectural complexity \cite{ref1}. To address these limitations, recent work has explored fluid antenna systems (FAS) \cite{ref21,ref22,ref23,ref24,ref25,ref26,ref27,ref29,ref30,ref31,ref32,ref33,ref34,ref35,ref36,ref361,ref362,ref37,ref38,ref39,ref40,ref41,ref42} to enhance spectral efficiency and reliability in large-scale MIMO for 6G networks \cite{ref22}.
Unlike fixed-position antenna (FPA) arrays with static element locations, FAS enables the radiating element or an equivalent port to be reconfigured within a prescribed aperture, allowing the terminal to sample the wireless channel at multiple spatial positions \cite{ref23}. Such reconfiguration can be realized through fluidic structures \cite{ref24}, reconfigurable metasurfaces \cite{ref25}, or spatially distributed antenna pixels \cite{ref26}, thereby providing additional spatial degrees of freedom that are largely decoupled from the RF-chain budget \cite{ref27}.

{Alongside FAS, the movable antenna (MA) paradigm represents a significant research branch in this area, with extensive recent efforts devoted to MIMO capacity characterization, channel modeling, and near-field optimizations \cite{ref42MA1,ref42MA2,ref42MA3,ref42MA4}. To harness the spatial degrees of freedom provided by these flexible antennas, existing literature has paired them with conventional multiuser transmission strategies, such as space-division multiple access (SDMA) \cite{ref30,ref38} and non-orthogonal multiple access (NOMA) \cite{ref37}. While such FAS-based schemes are capable of exploiting spatial channel variations, they still have the limitations of their underlying access mechanisms, i.e., the strong dependence of SDMA on accurate CSIT for effective interference cancellation and the potential loss of spatial multiplexing gain in multi-antenna NOMA \cite{ref6}. In contrast, RSMA offers a more flexible interference-management framework by jointly transmitting common and private streams, making its integration with FAS a distinct and more resilient solution.}

The unique capabilities of RSMA and FAS motivate the integration of these two technologies \cite{ref20,ref43,ref44,ref45,ref46,ref461,ref47}. RSMA offers flexible interference management and rate adaptation via message splitting, whereas FAS provides adaptive port reconfiguration that enables spatial sampling with limited RF hardware. Recent studies have begun to formalize FAS-RSMA designs: \cite{ref20} applies FAS-RSMA in the Non-Terrestrial Networks (NTNs) to maximize the user fairness; \cite{ref43} considers an FAS-aided RSMA framework for integrated sensing and communications (ISAC); \cite{ref44} develops a downlink FAS-RSMA architecture; \cite{ref45} examines STAR-RIS-assisted RSMA with FAS users; And \cite{ref46} analyzes UAV-relay-assisted RSMA networks with FAS terminals; And \cite{ref461} derived closed-form outage probability (OP) and ergodic rate expressions for multi-user single-input single-output (SISO) RSMA system. 

\begin{table*}[t]
\caption{Comparison of this work with existing FAS-RSMA studies.}
\label{tab:comparison}
\centering
\small
\setlength{\tabcolsep}{3pt}
\renewcommand{\arraystretch}{1.1}
\begin{tabular}{@{}c c c c c c c@{}}
\toprule
\textbf{Ref.} &
\begin{tabular}{@{}c@{}}\textbf{Antenna}\\ \textbf{Config.}\end{tabular} &
\begin{tabular}{@{}c@{}}\textbf{Performance}\\ \textbf{Analysis}\end{tabular} &
\begin{tabular}{@{}c@{}}\textbf{Precoder}\\ \textbf{Design}\end{tabular} &
\begin{tabular}{@{}c@{}}\textbf{Optimization}\end{tabular} &
\begin{tabular}{@{}c@{}}\textbf{User}\\ \textbf{Fairness}\end{tabular} &
\textbf{Key Distinction} \\
\midrule
{\cite{ref20}} & MISO & $\times$ & \checkmark & \checkmark & \checkmark & Max-min rate optimization without performance analysis. \\
{\cite{ref43}} & MISO & $\times$ & \checkmark & \checkmark & $\times$ & Secrecy rate optimization without performance analysis. \\
{\cite{ref44}} & SISO & \checkmark & $\times$ & $\times$ & $\times$ & OP analysis without closed-form expression. \\
{\cite{ref45}} & SISO & \checkmark & $\times$ & $\times$ & $\times$ & OP and ergodic rate analysis without closed-form expressions. \\
{\cite{ref46}} & SISO & \checkmark & $\times$ & $\times$ & $\times$ & OP analysis without closed-form expression. \\
{\cite{ref461}} & SISO & \checkmark & $\times$ & $\times$ & $\times$ & Closed-form OP and ergodic rate analysis in SISO setting. \\
\midrule
\textbf{This work} & \textbf{MISO} & \checkmark & \checkmark & \checkmark & \checkmark & \textbf{MISO FAS-RSMA with closed-form OP and optimization.} \\
\bottomrule
\end{tabular}
\end{table*}

Although recent studies have advanced the performance analysis of FAS-RSMA, a comprehensive analytical characterization of OP remains challenging. In multi-antenna downlink systems, the transmit precoder couples users’ effective channels, which complicates closed-form analysis \cite{ref47}. Meanwhile, the two-stage decoding structure of RSMA introduces coupled feasibility conditions between the common and private SINRs \cite{ref44}. The integration of FAS further increases the difficulty, since dynamic port reconfiguration operates over many spatial sampling points and they are spatially correlated \cite{ref32}. As a result, existing OP studies \cite{ref29,ref44,ref45,ref46,ref461} are mostly limited to SISO settings with fixed power allocation and typically rely on computationally intensive numerical evaluation rather than closed-form characterization, thereby leaving fairness-oriented designs such as min-max outage unexplored.

To bridge the gap above and tackle the foregoing challenges, we develop an analytical framework for FAS-RSMA that explicitly models spatial correlation among FAS ports in a multi-user multiple-input single-output (MISO) downlink employing RSMA. Spatial dependence across candidate ports is captured via block-correlation modeling~\cite{ref22,ref41,ref42}, yielding closed-form expressions for OP. In addition, to promote fairness across users, we adopt a min-max OP objective and design a resource-allocation scheme that balances reliability under practical constraints. {To explicitly highlight the novelty of this paper, we summarize the key differences between our work and existing FAS-RSMA studies in Table \ref{tab:comparison}.} The main contributions are summarized as follows:
\begin{itemize}
    \item \textbf{MISO FAS-RSMA Downlink Analysis:} We develop a tractable analysis framework for a multiuser MISO system employing RSMA and FAS. To the best of our knowledge, this work represents one of the earliest systematic studies of FAS performance in a multiuser MISO downlink under spatially correlated ports.

    \item \textbf{Closed-form OP Derivations:} Within a consistent analytical framework, we derive closed-form expressions for users’ OP by adopting zero-forcing (ZF) precoding and maximum-ratio transmission (MRT) precoding. In contrast to conventional semi-analytical expressions with numerical approximation, the proposed closed-form expressions enable fast and accurate performance assessment without high-dimensional integration. The closed-form OP expressions are also applicable to the scenario where users are equipped with uncorrelated antenna ports.

    \item \textbf{Min-max OP Optimization:} Leveraging the closed-form OP characterization, we formulate a fairness-oriented min-max OP optimization problem and develop an efficient solver. Specifically, a bisection on the target OP is performed, where each feasibility check reduces to a linear program with a closed-form power-allocation solution. This yields the proposed \emph{bisection with closed-form optimal power allocation (BCOPA)} algorithm, enabling fast and systematic resource allocation across users while achieving the optimal min-max OP value.

    \item \textbf{Extensive Numerical Validation:} Using the closed-form OP expressions, we validate our analysis against Monte Carlo simulations. The analytical results closely match the simulations, and FAS-RSMA consistently achieves lower OP than the fixed-antenna baseline by up to $ 92\%$ over the tested signal-to-noise ratio (SNR) range. We also evaluate the proposed BCOPA algorithm, which attains the same optimal min-max OP as exhaustive search (ES) but with significantly lower computational cost (runtime drop by approximately $99\%$), demonstrating effective power allocation across users. Relative to fixed power splits, min-max optimization yields an approximate $1$\,dB SNR gain at a fixed outage target for the worst user, demonstrating improved fairness and reliability.
\end{itemize}

\textit{Notation:} Bold uppercase letters (e.g., $\mathbf{A}$) denote matrices, bold lowercase letters (e.g., $\mathbf{a}$) denote column vectors, and standard letters denote scalars. All the mathematical notations and parameter definitions are listed in Table \ref{table1}. 

The remainder of this paper is organized as follows. Section~II details the multiuser MISO downlink with the FAS-RSMA system model and the adopted block-correlation model. Section~III develops the analytical framework and derives closed-form OP expressions. Section~IV formulates the min--max OP design and presents the solver. Section~V reports numerical results across diverse system configurations and channel settings. Finally, Section~VI concludes the paper and outlines directions for future research. 

\begin{table}[ht]
\center
\caption{Mathematical notations and parameter definitions.}
\label{table1}
\vspace{+1mm}
\begin{tabular}{@{}l@{\hspace{5mm}}l@{}}
\toprule
\textbf{Symbol}&\textbf{Meaning}\\
  \midrule
  \multicolumn{2}{c}{\textbf{Mathematical Notations}} \\
  \midrule
  $f_X(x)$& Probability density function (PDF) of a variable $x$\\
  $F_X(x)$& Cumulative distribution function (CDF) of a variable $x$\\
  ${\left(  \cdot  \right)^H}$&Conjugate transpose\\
  $\left\| {\mathbf{x}} \right\|$&2-norm of a vector ${\mathbf{x}}$\\
  $\mathbb{E}\{\cdot\}$ &Expectation operator\\
  $\mathrm{Pr}\left(  \cdot  \right)$ &Probability operator\\
  ${x_i}$&The $i$-th element of a vector ${\mathbf{x}}$\\
  $\prod\limits_{d=1}^{D} x_d$&The multiplication of $x_d$ from $d=1$ to $D$\\
  $\sum\limits_{u = 1}^U {{x_u}} $&The sum of $x_u$ from $u=1$ to $U$\\
  ${{\mathbf{X}}^{ - 1}}$&Matrix inverse \\
  $d f/d x$&Derivative of $f$ with respect to $x$\\
  $\int_{a}^{b} dx$&Integration from $a$ to $b$ with respect to $x$\\
  $\log x$&Logarithm of a scalar $x$ with base 2\\
  $J_{0}$&Zeroth-order Bessel function of the first kind\\ 
  $\mathcal{O}$&Order of computational complexity\\
  $\Gamma \left(\alpha,\theta \right)$ &Gamma distribution with shape $\alpha$ and scale $\theta$\\
  \multirow{2}{*}{$\mathrm{Naka} \left(m, \Omega_u\right)$}&{Nakagami-$m$ distribution with}\\&{ fading figure $m$ and average power $\Omega_u$}\\
  \multirow{2}{*}{$\chi'^2_{2\alpha}(\lambda)$}&{Scaled noncentral chi-square with }\\&{ $2\alpha$ degrees of freedom and noncentrality $\lambda$}\\
  \multirow{2}{*}{$ \mathcal{C}\mathcal{N}\left( {a,{{\sigma^2 }}} \right)$}&{Complex Gaussian distribution with}\\&{ mean $a$ and variance $\sigma^2 $}\\
  \midrule
  \multicolumn{2}{c}{\textbf{Parameter Definitions}} \\
  \midrule
  $L$& Number of transmit antennas at the BS\\
  $U$& Number of users\\
  $D$& Number of sub-blocks \\
  $N$ & Number of antenna ports \\
  $L_{d}$ & Size of $d$-th block \\
  $\rho_d$ & Spatial correlation coefficient of $d$-th block \\
  $\alpha=Lm$ & Equivalent Gamma shape parameter \\
  $\kappa_d= \frac{2 \rho_d}{1-\rho_d}$ & Equivalent correlation coefficient of $d$-th block \\
  $\gamma^{u}_{\mathrm{th}}$ & Overall outage threshold for user $u$ \\
  $P$ & Effective transmit-power after normalization \\
  $t_c$ & Power allocation coefficient of the common stream \\
  $t_u$ & Power allocation coefficient of the $u$-th private stream \\
  $Q(x)$&$\int_x^\infty \frac{1}{\sqrt{2\pi}}e^{-\frac{1}{2}t^2}dt$\\
  \bottomrule
\end{tabular}
\label{tbl}
\end{table}

\section{System model for FAS-RSMA Communication}
\begin{figure*}
    \setlength{\belowcaptionskip}{-0.9cm}
    \centering
    \includegraphics[width=0.9\textwidth]{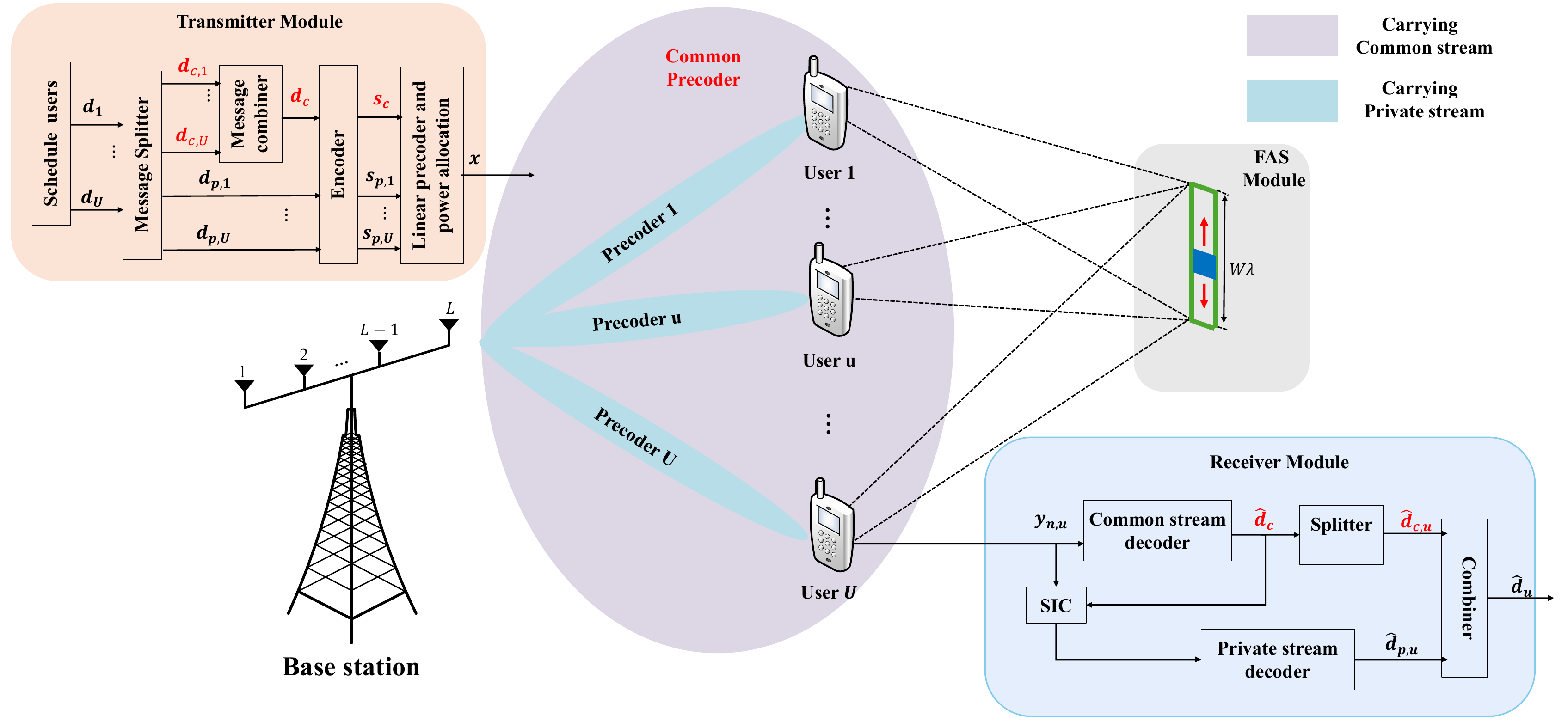}
    \caption{The proposed multiuser MISO downlink FAS-RSMA system framework.}
    \label{fig:System model}
\end{figure*}

We consider a downlink communication system where a BS equipped $L$ antennas serves $U$ users. {All users are equipped with one-dimensional linear fluid antennas, where a single active antenna (i.e., a single radio-frequency chain) enables switching among $N$ discrete ports.} These ports are distributed within a linear space of length $W \lambda$, where $W$ is a normalization parameter and $\lambda$ is the carrier wavelength\footnote{{Practically, FAS deployment in space-constrained terminals relies on either microfluidic architectures or solid-state RF-pixel designs \cite{ref1,ref22}. To manage the pilot overhead associated with dense virtual ports, user terminals typically employ fast port selection algorithms, such as compressive sensing or learning-assisted sequential probing, during the training phase \cite{ref1,ref35}.}}. The overall system framework is shown in Fig.~\ref{fig:System model}. {The use of the same FAS configuration across users corresponds to a homogeneous multiuser setting, e.g., same-type terminals in hotspot or massive-connectivity scenarios \cite{ref1,ref22,ref27}, and is adopted here for analytical tractability and fair comparison.} In this section, we detail the signal model at the transmitter and receiver, the performance metrics, and the receiver-side channel-correlation model. 

\subsection{Signal Model}
{At the BS, the intended data message for each user $u$, denoted by $d_u$, is split into a common part $d_{c,u}$ and a private part $d_{p,u}$. All common-part messages $\{d_{c,1}, \dots, d_{c,U}\}$ are then grouped and encoded into a single common stream $s_{c}$, while the private messages $d_{p,u}$ are encoded independently into dedicated private streams, denoted as $s_{p,u}$.} {We assume $\mathbb{E}\{|s_c|^2\}=1$ and $\mathbb{E}\{|s_{p,u}|^2\}=1$, and the power allocation factors satisfy $t_c\ge 0$, $t_u\ge 0$, and $t_c+\sum_{u=1}^{U} t_u = 1$. Following the 1-layer RS strategy of RSMA \cite{ref6,ref7}, the overall transmitted signal at the BS can be expressed as
\begin{equation}
  \mathbf{x}= \sqrt{P}  \left ( \sqrt{t_c} \mathbf{w}_c {s}_{c}  +  \sum_{u=1}^{U} \sqrt{t_u} \mathbf{w}_{n,u} {s}_{p,u} \right ), 
\label{RSMA_transmit_signal} 
\end{equation}
where $t_c$ and $t_u$ denote the power allocation factors for the common and private data streams, respectively. The overall transmit-power budget is imposed at the superimposed-signal level through the scaling $\sqrt{P}$ together with the coefficients $t_c$ and $t_u$. Specifically, under a long-term average total transmit-power constraint, if $P_0$ denotes the total transmit-power budget, the equivalent normalization factor of the superimposed transmit signal is given by
\begin{equation}
\eta=t_c\,\mathbb{E}\{\|\mathbf{w}_c\|^2\}+\sum_{u=1}^{U} t_u\,\mathbb{E}\{\|\mathbf{w}_{n,u}\|^2\}.
\end{equation}
Accordingly, we define $P=P_0/\eta$. For notational simplicity, this normalization factor is absorbed into $P$, and hence $P$ should be interpreted as the effective transmit-power parameter after overall normalization. Meanwhile, $\mathbf{w}_c \in \mathbb{C}^{L \times 1}$ and $\mathbf{w}_{n,u} \in \mathbb{C}^{L \times 1}$ represent the precoding vectors for the common and the $u$-th private data stream at the $n$-th port, respectively. The precoding vectors are designed according to a hybrid ZF-MRT principle\footnote{{RSMA and conventional linear precoding are not mutually exclusive. Rather, RSMA provides a more general framework that can incorporate ZF or MMSE designs \cite{ref3}. Under favorable conditions, RSMA reduces to conventional SDMA, whereas its main advantage appears in challenging scenarios such as imperfect CSI or non-orthogonal user channels \cite{ref6,ref7}.}}, shown as
\begin{equation}
  \mathbf{w}_{n,u}=\hat{\mathbf{w}}_{n,u} \left \|\mathbf{h}_{n,u}  \right \|, \quad \mathbf{w}_c=\sum_{u \in \mathcal{U}} \mathbf{w}_{n,u},
\label{precoding} 
\end{equation}
where $\mathbf{h}_{n,u}=\left[h^1_{n,u}, \dots, h^L_{n,u} \right]^T \in \mathbb{C}^{L \times 1}$ denotes the channel vector between the BS and the $n$-th port of user $u$. Let $\mathbf{H}_n=\left[\mathbf{h}_{n,1}, \dots,\mathbf{h}_{n,U} \right] \in \mathbb{C}^{L \times U}$ denote the aggregated channel matrix. Then, $\hat{\mathbf{w}}_{n,u} \in \mathbb{C}^{L \times 1}$ is defined as the $u$-th column vector of the zero-forcing precoding matrix $\mathbf{W} = \mathbf{H}_n (\mathbf{H}_n^H \mathbf{H}_n)^{-1} \in \mathbb{C}^{L \times U}$.}
Based on the ZF-MRT design with perfect CSI, the received signal can be expressed as
\begin{align}
y_{n,u} & =\sqrt{P}\left(\mathbf{h}^{H}_{n,u} \mathbf{w}_{c}\sqrt{t_{c}} s_{c} +\sum_{j \in \mathcal{U}}\mathbf{h}^{H}_{n,u} \mathbf{w}_{n,j} \sqrt{t_{j}} s_{p,j} \right)+w_{u} \notag\\
& =\sqrt{P}\left(\sqrt{t_{c}} s_{c}+\sqrt{t_{u}} s_{p, u}\right)\zeta_{n,u} +w_{u},
\label{RSMA_recevied_signal} 
\end{align}
where $\zeta_{n,u}=\mathbf{h}^{H}_{n,u} \mathbf{w}_{n,u}= \left \|\mathbf{h}_{n,u}  \right \|$ denotes the effective channel coefficient at the $n$-th port, and $w_{u} \sim  \mathcal{C} \mathcal{N}(0,\sigma^{2})$ is the complex additive white Gaussian noise (AWGN). We assume that the channel follows a Nakagami-$m$ distribution\footnote{{The Nakagami-$m$ model differs from the field-response model \cite{ref42MA4}, which provides a physics-based spatial description of the FAS aperture. In this work, the Nakagami-$m$ distribution is adopted as a flexible and analytically tractable model for small-scale fading, as it captures varying fading severities and yields a Gamma-distributed power gain suitable for outage analysis.}}, denoted by $h^l_{n,u} \sim \mathrm{Naka} \left(m, \Omega_u\right)$, where $m$ and $\Omega_u$ are the fading figure and average power parameters, respectively. Following the FAS concept, we assume that the optimal port with the maximum overall gain is activated, written as
\begin{equation}
    n^{*}=\arg \max _{n \in[1, N]} \left |\zeta_{n,u}  \right |^2 ,
    \label{select}
\end{equation}
where $\left |\zeta_{ n^{*},u}  \right |^2$ denotes the maximum overall gain of the selected port. For notational simplicity, we henceforth suppress the index $n^{*}$ and define $\left |\zeta_{u}  \right |^2 \triangleq \left |\zeta_{ n^{*},u}  \right |^2$. {We assume that the optimal port specific to each individual user $u$ is selected independently within their own fluid antenna aperture to maximize their respective effective channel gain.}

{Based on the RSMA decoding procedure, the decoding process for all users consists of two stages. In the first stage, the common stream is decoded while treating other signals as noise. After the common stream is successfully decoded, it is removed from the received signal via SIC. In the second stage, each user decodes its intended private stream from the residual signal while treating any remaining interference as noise. The recovered common and private parts are then combined to reconstruct the original message intended for each user.} The expressions for the common SINR $\Gamma_{c,u}$ and private SINR $\Gamma_{p,u}$ are given by
\begin{subequations}\label{SINRs}
\begin{align}
 &\Gamma_{c,u}=\frac{\bar{\gamma } t_c \left |\zeta_{u}  \right |^2     }{\bar{\gamma } t_u \left |\zeta_{u}  \right |^2  +1 }, \\
 &\Gamma_{p,u}=\bar{\gamma } t_u  \left |\zeta_{u}  \right |^2,
\end{align}
\end{subequations}
with $\bar{\gamma }=\frac{P}{\sigma^{2}}$ represents the average transmit SNR. Based on the (\ref{SINRs}), we can define the achievable common and private rates as
\begin{subequations}\label{rates}
\begin{align}
 & R_{c,u}=\log \left ( 1+\Gamma_{c,u} \right), \\
 &R_{p,u}=\log \left ( 1+\Gamma_{p,u} \right).
\end{align}
\end{subequations}

\subsection{Performance Metrics Model}
{The OP is a fundamental reliability performance metric in wireless communication systems that quantifies the likelihood of a user failing to decode its intended message. Since decoding of the RSMA receiver is performed through a successive two-stage procedure, an outage event for user $u$ occurs under two mutually exclusive conditions: (a) failure to decode the common stream, or (b) successful decoding of the common stream but failure to decode the private stream. Let $\mathcal{E}_{c} \triangleq \{R_{c,u} > \tilde{R}_c\}$ and $\mathcal{E}_{p} \triangleq \{R_{p,u} > \tilde{R}_{p,u}\}$ denote the events of successful decoding for the common and private streams with target rates $\widetilde{R}_{c} $ and $\widetilde{R}_{p,u} $, respectively. The overall sequential outage event can be logically expressed as $\mathcal{E}_{c}^f \cup (\mathcal{E}_{c} \cap \mathcal{E}_{p}^f)$, where the superscript $f$ denotes the complement (failure event). According to elementary set theory and Boolean algebra, this union simplifies to $\mathcal{E}_{c}^f \cup \mathcal{E}_{p}^f$. Consequently, the OP of user $u$ is given by \cite{ref11,ref461}
\begin{align} P_{out}^{u} =& \mathrm{Pr}(\mathcal{E}_{c}^f \cup \mathcal{E}_{p}^f)  \notag\\ 
=& \mathrm{Pr}\left(\frac{\bar{\gamma}t_{c}|\zeta_{u}|^{2}}{\bar{\gamma}t_{u}|\zeta_{u}|^{2}+1} \le \gamma_{th}^{c,u} \cup \bar{\gamma}t_{u}|\zeta_{u}|^{2} \le \gamma_{th}^{p,u}\right) \notag\\ 
=& \mathrm{Pr} \left (\left |\zeta_{u}  \right |^2 \le\tilde{\gamma}^{c,u}_{\mathrm{th}}\cup \left |\zeta_{u}  \right |^2 \le\tilde{\gamma}^{p,u}_{\mathrm{th}}  \right ) \notag\\
=& \mathrm{Pr}\left ( \left |\zeta_{u}  \right |^2\le \max \left \{ \tilde{\gamma}^{c,u}_{\mathrm{th}},\tilde{\gamma}^{p,u}_{\mathrm{th}} \right \}  \right ) \notag\\
=& F_{\left |\zeta_{u}  \right |^2} \left (\gamma^{u}_{\mathrm{th}}  \right ),
\end{align}}where the SINR thresholds for the common and private streams are $\gamma^{c,u}_{\mathrm{th}}=2^{\widetilde{R}_{c}}-1$ and $\gamma^{p,u}_{\mathrm{th}}=2^{\widetilde{R}_{p,u}}-1$, respectively, and the per-user requirement is $\gamma^{u}_{\mathrm{th}} = \max \left \{ \tilde{\gamma}^{c,u}_{\mathrm{th}},\tilde{\gamma}^{p,u}_{\mathrm{th}} \right \}$ with $\tilde{\gamma}^{c,u}_{\mathrm{th}}$ and $\tilde{\gamma}^{p,u}_{\mathrm{th}}$ defined as
\begin{subequations}\label{gamma}
\begin{align}
 &\tilde{\gamma}^{c,u}_{\mathrm{th}}=\frac{\gamma^{c,u}_{\mathrm{th}}}{\bar{\gamma } \left (t _c-t_{u} \gamma^{c,u}_{\mathrm{th}}  \right ) } 
    , \\
 &\tilde{\gamma}^{p,u}_{\mathrm{th}}=\frac{\gamma^{p,u}_{\mathrm{th}}}{\bar{\gamma } t_u  }.
\end{align}
\end{subequations}
{The ordering between $\tilde{\gamma}_{\mathrm{th}}^{c,u}$ and $\tilde{\gamma}_{\mathrm{th}}^{p,u}$ is not predetermined. Rather, it is jointly determined by the SINR thresholds $\{\gamma_{\mathrm{th}}^{c,u},\gamma_{\mathrm{th}}^{p,u}\}$ and the power-allocation coefficients $\{t_c,t_u\}$. While the common data rate is globally limited by the weakest user in the system, the individual user OP is determined by the specific SINR requirements of both streams.} \\
\textbf{Remark 1:} \textit{To ensure that (\ref{gamma}) is valid and non-negative, the thresholds must satisfy 
\begin{equation}
{\gamma}^{c,u}_{\mathrm{th}}<\frac{t_{c}}{ t_{u}}, \forall u,
\end{equation}
which means that if thresholds exceed these bounds, the algebraic SINR in \eqref{gamma} becomes nonphysical (negative or undefined), which in turn makes the decoding constraints infeasible. In this case, the outage event occurs with probability one, and the analytical
expressions lose validity.}

\subsection{Channel Correlation Model}
{Since adjacent ports are closely spaced along the fluid antenna aperture, the underlying complex channel coefficients observed at different ports exhibit non-negligible spatial correlation, which must be accounted for in the performance analysis. We characterize this dependence through the covariance matrix $\bm{\Sigma}_{l,u}\in\mathbb{R}^{N\times N}=\mathbb{E}\!\left\{\mathbf{h}^{(l)}_{u}\left(\mathbf{h}^{(l)}_{u}\right)^{H}\right\} $, where $\mathbf{h}^{(l)}_{u}\triangleq [h^l_{1,u},\ldots,h^l_{N,u}]^{T}$ collects the channel coefficients across the $N$ ports of $l$-th transmitted antenna branch. According to Jakes' model \cite{ref21,ref22}, the spatial correlation coefficient between the $k$-th and $q$-th ports can be modeled as
\begin{equation}
    \sigma_{k, q}=J_{0}\left(\frac{2 \pi|k-q| W}{N-1}\right),
    \label{Jakes}
\end{equation}
where $J_{0}$ is the zeroth-order Bessel function of the first kind. Consequently, $\bm{\Sigma}_{l,u}$ admits a Toeplitz form shown as 
\begin{equation}
\begin{aligned}
\boldsymbol{\Sigma}_{l,u} \in \mathbb{R}^{N \times N} & =\textbf{toeplitz}\left(\sigma_{1,1}, \sigma_{1,2}, \cdots, \sigma_{1, N}\right) \\
& =\left(\begin{array}{cccc}
\sigma_{1,1} & \sigma_{1,2} & \cdots & \sigma_{1, N} \\
\sigma_{1,2} & \sigma_{1,1} & \cdots & \sigma_{1, N-1} \\
\vdots & & \ddots & \vdots \\
\sigma_{1, N} & \sigma_{1, N-1} & \cdots & \sigma_{1,1}
\end{array}\right),
\end{aligned}
\label{eq:toeplitz_matrix}
\end{equation}
which is Hermitian with entries determined solely by the relative port separation. While this structure captures the spatial correlation along the aperture, the resulting non-diagonal covariance complicates exact distributional analysis. In particular, closed-form expressions for key performance metrics are generally hard to obtain.}

\section{Performance analysis for FAS-RSMA}
In this section, we analyze the performance of FAS-RSMA system with emphasis on the effect of spatial correlation on OP. The block-correlation model is employed to effectively capture and characterize the intricate spatial correlation patterns observed along the fluid-antenna aperture. Building on this characterization, we derive closed-form expressions for the OP.

\subsection{Block-Diagonal Matrix Approximation}
To manage the analytical difficulty posed by the Toeplitz covariance, we adopt a block-correlation approximation \cite{ref22}, in which the set of correlated port coefficients is partitioned into several independent blocks. Under this representation, the spatial covariance of user $u$ is approximated by a block-diagonal matrix shown as
\begin{equation}
\widehat{\boldsymbol{\Sigma}}_{l,u} \in \mathbb{R}^{N \times N} =
\left[
\begin{array}{cccc}
\mathbf{A}_1 & \mathbf{0} & \cdots & \mathbf{0} \\
\mathbf{0}   & \mathbf{A}_2 & \cdots & \mathbf{0} \\
\vdots       & \vdots       & \ddots & \vdots \\
\mathbf{0}   & \mathbf{0}   & \cdots & \mathbf{A}_D
\end{array}
\right],
\label{eq:block_matrix}
\end{equation}
where each block matrix $\mathbf{A}_d \in \mathbb{R}^{L_d \times L_d}$ for $d=1, \ldots, D$ is given by
\begin{equation}
\mathbf{A}_d \in \mathbb{R}^{L_d \times L_d} =
\left[
\begin{array}{cccc}
1     & \rho_d & \cdots & \rho_d \\
\rho_d & 1     & \cdots & \rho_d \\
\vdots & \vdots & \ddots & \vdots \\
\rho_d & \rho_d & \cdots & 1
\end{array}
\right].
\label{eq:block_matrix_single}
\end{equation}
Here, $D$ denotes the number of sub-blocks, determined as the count of eigenvalues exceeding unity, with sizes $\{L_d\}_{d=1}^{D}$ satisfying $\sum_{d=1}^{D} L_d=N$. Each block is modeled with a common correlation coefficient $\rho_d \in \left [ 0,1  \right ]$. The block-correlation approximation thus partitions the original correlated ports into $D$ independent blocks, where each block shares a common correlation coefficient. To fit the approximation to the target covariance $\boldsymbol{\Sigma}_{l,u}$, the parameters $D$, $\left \{\rho_d  \right \}_{d=1}^{D}$, and $\left \{L_d  \right \}_{d=1}^{D}$ should be optimized by minimizing the following objective function:
\begin{equation}
\arg \min_{D, \left \{\rho_d  \right \}_{d=1}^{D}, \left \{L_d  \right \}_{d=1}^{D}} \; \mathrm{dist}\big( \boldsymbol{\Sigma}_{l,u}, \widehat{\boldsymbol{\Sigma}}_{l,u} \big),
\label{eq:optimization}
\end{equation}
{where $\mathrm{dist}\big( \cdot, \cdot \big)$ denotes the Euclidean distance between the eigenvalues of the two matrices, defined as
\begin{equation}
\mathrm{dist}\big( \boldsymbol{\Sigma}_{l,u}, \widehat{\boldsymbol{\Sigma}}_{l,u} \big) 
= \big\| \mathrm{Eig}(\boldsymbol{\Sigma}_{l,u}) - \mathrm{Eig}(\widehat{\boldsymbol{\Sigma}}_{l,u}) \big\|^2,
\label{eq:distance_metric}
\end{equation}
with $\mathrm{Eig} \left( \cdot \right )$ representing the vector of eigenvalues of the matrix. The algorithmic procedure for solving \eqref{eq:optimization} is detailed in \cite{ref41,ref461}. The main idea is to transform the block-correlation optimization into a low-complexity eigenvalue-allocation problem, where a greedy sequential heuristic is used to distribute the non-dominant eigenvalues among the individual blocks.} 

\subsection{Derivation of Outage Probability}
In this section, we present the detailed derivation of a closed-form expression for the OP of the FAS-RSMA system. Since $h^l_{n,u} \sim \mathrm{Naka} \left(m, \Omega_u\right)$, the $|h^l_{n,u}|^2$ can be denoted as Gamma random variable, i.e., $|h^l_{n,u}|^2 \sim \Gamma \left(m,\theta \right)$, where $\theta=\Omega_u /m$. Hence, we can have the following proposition:\\
\textbf{Proposition 1:} \textit{{For a fixed receive port $n$, suppose the channel power gain satisfy $|h^{\,l}_{n,u}|^{2}\sim \Gamma(m,\theta)$ and are independent and identically distributed (i.i.d.) for $l=1,\dots,L$.} If there is only single receive port, we can define the distribution of the effective channel power $\left|\zeta_{u}  \right |^2 $ as 
\begin{equation}
\left |\zeta_{u}  \right |^2=\sum_{l=1}^L |h^l_{n,u}|^2 \sim \Gamma(\alpha,\theta),\quad \alpha=L m.
\label{eq:17}
\end{equation}
}

\textit{Proof:} Since $|\zeta_u|^2$ is the sum of $L$ i.i.d. Gamma random variables with the same scale parameter $\theta$, it follows from the additive property of the Gamma distribution that $|\zeta_u|^2 \sim \Gamma(Lm,\theta)$.

{To capture port-level spatial correlation, we employ a one-factor construction within each sub-block $d$. We introduce auxiliary complex Gaussian layers indexed by $q=1,\dots,\alpha$. Specifically, let $\hat{h}_{i,q}^{d}$ denote the $q$-th auxiliary Gaussian component for the $i$-th port in sub-block $d$. We can formulate $\hat{h}_{i,q}^{d}$ based on the one-factor model as follows}: 
\begin{equation}\label{eq:one-factor}
\hat{h}^{d}_{i,q}=\sqrt{\theta}\Big(\sqrt{1-\rho_d}\,z_{i,q}+\sqrt{\rho_d}\,z_{0,q}\Big),
\end{equation}
where $z_{i,q}, z_{0,q}\overset{\text{i.i.d.}}{\sim}\mathcal{C} \mathcal{N}(0,1)$ and $\rho_d$ denotes the within-block correlation parameter obtained from the block-diagonal matrix approximation. In \eqref{eq:one-factor}, the contribution scaled by $\sqrt{1-\rho_d}$ models the independent component, whereas the term scaled by $\sqrt{\rho_d}$ acts as a common factor shared by all coefficients in block $d$. Then, we define the per-port power in block $d$ by aggregating the layer contributions as
\begin{equation}\label{eq:Xid-sum}
X^d_i=\sum_{q=1}^{\alpha}\big|\hat{h}^{d}_{i,q}\big|^2.
\end{equation}
For $\alpha\in\mathbb{N}$, \eqref{eq:Xid-sum} is a finite sum. For real $\alpha>0$, we adopt a standard conditional construction that preserves both the marginal distribution (defined in \textbf{Proposition 1}) and joint distribution.

We introduce the parameters
\begin{equation}\label{eq:sigma-kappa}
\sigma_d^2\triangleq \theta(1-\rho_d),\qquad \kappa_d\triangleq \frac{2\rho_d}{1-\rho_d},
\end{equation}
and define the common mixing variable
\begin{equation}\label{eq:Ud-def}
U_d\triangleq \sum_{q=1}^{\alpha}|z_{0,q}|^2\ \sim\ \Gamma(\alpha,1).
\end{equation}
Conditioned on $U_d=u$, the factorization in \eqref{eq:one-factor} yields
\begin{equation}\label{eq:aa}
 \hat{h}^{d}_{{i,q} \mid U_d=v} \sim\ \mathcal{C} \mathcal{N}(\mu_q,\sigma_d^2), 
\end{equation}
where the mean and variance parameters are denoted as $\mu_q=\sqrt{\theta\rho_d}\,z_{0,q}$ and $\sigma_d^2=\theta \left( 1-\rho_d \right)$, respectively.
Based on the above results, we have the following proposition:\\
\textbf{Proposition 2:} \textit{For each port $i$ in block $d$, the conditional distribution of $X^{d}_{i}$ given $U_d=v$ is
\begin{equation}\label{eq:cond-ncx2}
X^d_{i \mid U_d=v} = \sum_{q=1}^{\alpha}\big|\hat{h}^{d}_{i,q}\big|^2  \stackrel{d}{=} \frac{\sigma_d^2}{2}\,\chi'^2_{2\alpha}\!\Big(\lambda=\kappa_d v\Big) 
\end{equation}
where $\chi'^2_{2\alpha}\!\Big(\lambda=\kappa_d v\Big)$ represents a scaled noncentral chi-square with $2\alpha$ degrees of freedom and noncentrality $\lambda=\kappa_d v$, and $\kappa_d$ is defined as
\begin{equation}
 \kappa_d\triangleq \frac{2\rho_d}{1-\rho_d}.   
 \label{eq:24}
\end{equation}
}

\textit{Proof:} See Appendix A.\\
\textbf{Remark 2:}\textit{ Based on the Equation~\eqref{eq:cond-ncx2} and $U_d\sim\Gamma(\alpha,1)$, we can have the following Laplace transform of the \emph{marginal} distribution for each port:
\begin{align}\label{eq:LT-marginal}
\mathcal{L}_{X^d_i|U}(t) &= (1+\sigma_d^2 t)^{-\alpha}\exp\!\left(-\frac{\kappa_d v\,\sigma_d^2 t}{2(1+\sigma_d^2 t)}\right),\notag\\
\mathcal{L}_{X^d_i}(t) &= \mathbb{E}_U[\mathcal{L}_{X^d_i|U}(t)]
= (1+\theta t)^{-\alpha},
\end{align}
which is the Laplace transform of $\Gamma(\alpha,\theta)$ \emph{for any real $\alpha>0$}. Thus, the introduced common random variable representation is valid without requiring an integer $\alpha$.
}

{Based on the conditional distribution established in \textbf{Proposition 2}, we can now derive the exact integral expression for the overall OP of the FAS-RSMA system, which is summarized in the following proposition.\\
\textbf{Proposition 3:} \textit{For the proposed multiuser MISO FAS-RSMA downlink, the exact OP of user $u$ under the block-correlation model is given by \begin{align}\label{eq:OP-final} P_{\mathrm{out}}^{u} = &\prod_{d=1}^{D}\ \int_{0}^{\infty} \frac{v^{\alpha-1}e^{-v}}{\Gamma(\alpha)} \notag\\  & \times \Big[1-Q_{\alpha}\!\Big(\sqrt{\kappa_d v},\sqrt{\frac{2 \gamma^u_{\mathrm{th}}}{\theta(1-\rho_d)}}\Big)\Big]^{L_d} dv, \end{align}
where $Q_{\alpha}(\cdot,\cdot)$ denotes the generalized Marcum $Q$-function.}}

{\textit{Proof:} See Appendix B.}

Based on \textbf{Proposition 3}, we can now state the following theorem.\\
\textbf{Theorem 1:} \textit{The closed-form expression\footnote{{In this paper, the term ``closed-form'' is used in the conventional wireless-communications sense, where finite expressions involving standard special functions are regarded as closed-form, provided that they can be efficiently evaluated using built-in functions in standard software without requiring numerical integration.}} for the OP of the $u$-th user in the proposed FAS-RSMA system can be represented as
\begin{align}
\widetilde{P}_{\mathrm{out}}^{u}\approx& \prod_{d=1}^{D}
\frac{1}{\Gamma(\alpha)}
\sum_{m=1}^{M_{\text{GL}}}
w_{m}^{(\alpha-1)} \notag\\
& \times\!\! \Bigg[1-Q_{\alpha}\!\Bigg(\sqrt{\kappa_d \,\xi_m},\sqrt{\frac{2\,\gamma_{\mathrm{th}}^{u}}{\theta(1-\rho_d)}}\Bigg)\Bigg]^{L_d},
\label{eq:theorem1}
\end{align}
where all the parameters are defined in Table \ref{table1}. And $w_m^{(\alpha-1)}$ is the weight of Gauss-Laguerre polynomials defined as
\begin{equation}
    w_{m}^{(\alpha-1)}=\frac{\Gamma(M_{\text{GL}}+\alpha) \xi_m}{M_{\text{GL}}!(M_{\text{GL}}+1)^{2}\left[L_{M_{\text{GL}}+1}^{(\alpha-1)}\left(\xi_m\right)\right]^{2}} .
\end{equation}
with $M_{\text{GL}}$ denoting the order of the Gauss-Laguerre polynomials and $L_{M_{\text{GL}}+1}^{(\alpha-1)}\left(\xi_m\right)$ being generalized Laguerre polynomials. 
}

\textit{Proof:} The proof of \textbf{Theorem 1} follows the generalized Gauss-Laguerre quadrature approach, which converts the integral expression into a finite weighted summation. For brevity, the detailed derivation is omitted here and can be found in \cite{ref461}.\\
\textbf{Remark 3:} \textit{The above theorem is derived under the assumption $\rho_d \in \left[0,1 \right)$. At the boundary $\rho_d=1$, the within-block covariance becomes rank-one and all ports in block $d$ are statistically indistinguishable, which coincides with the conventional FPA baseline. Based on \textbf{Proposition 1}, the corresponding OP is given by the CDF of the Gamma distribution shown as
\begin{equation}\label{tas}
\tilde{P}_{out}^{u}\!= \!\frac{\gamma\!\left(\alpha,\,\gamma_{\mathrm{th}}^{u}/\theta\right)}{\Gamma(\alpha)}\!= \!1\!-\!\exp\!\!\left(\!-\frac{\gamma_{\mathrm{th}}^{u}}{\theta}\right)
\sum_{n=0}^{\alpha-1}\frac{1}{n!}\left(\frac{\gamma_{\mathrm{th}}^{u}}{\theta}\right)^{n}.  
\end{equation}
where $\gamma(\cdot,\cdot)$ is the lower incomplete gamma function and $\gamma(\alpha,x)/\Gamma(\alpha)$ is the regularized form.
}

The derived OP expression has a block-wise product structure, with each sub-block CDF expressed as a finite weighted sum over Gauss-Laguerre nodes. Accordingly, the OP is determined by the SINR threshold, the SNR, and the number of antenna ports. For a fixed threshold, the OP decreases as the SNR and $N$ increase, which reflects the selection-diversity gain provided by port reconfiguration.

\section{Performance Optimization of FAS-RSMA}
{Motivated by balanced reliability, this section studies fairness-oriented allocation of common and private power in the FAS-RSMA downlink. Specifically, fairness is enforced by mathematically penalizing the worst-case user performance, ensuring that the optimal power allocation effectively balances the reliability across all scheduled users and minimizes severe performance disparities. Using the closed-form outage characterization from \textbf{Theorem 1}, we propose a min-max optimization problem that minimizes the maximum per-user OP subject to standard decoding-feasibility and sum-power constraints. We then outline an efficient solution procedure for this optimization.}
\subsection{Problem Formulation}
To improve user performance in FAS-RSMA under a fairness objective, we propose the following min-max optimization problem:
\begin{subequations}
\label{p}
\begin{align}
\mathop {\min} \limits_{\left\{t_c,\mathbf{t}\right\}}  \mathop {\max }\limits_{u \in \mathcal{U}  } \quad  & \widetilde{P}^{u}_{\mathrm{out}} \label{p1} \\
 \text { s.t. }\;\;\;\; &  t_{u}{\gamma}^{c,u}_{\mathrm{th}}<t_{c}, \forall u,
\label{p2}  \\
&t_c+\sum_{u=1}^{U} t_u \leq 1, \label{p3} 
\end{align}        
\end{subequations}
where (\ref{p2}) comes from the feasible decoding requirement in Remark 1, and (\ref{p3}) is the power-budget constraint. The above problem is a non-convex, non-smooth min-max program. The user threshold $\gamma^u_{\mathrm{th}}$ contains the maximum operation of reciprocal mappings in optimization variables, and the constraint (\ref{p2}) tightly couples variables. Moreover, the objective function $\widetilde{P}^{u}_{\mathrm{out}}$ arises from generalized Gauss-Laguerre quadrature of Marcum-Q terms, yielding complex mappings across users, which makes the problem difficult to solve. 

\subsection{Proposed Solution}
In this section, we present the proposed algorithm for solving problem (\ref{p}). To make the problem more tractable, we first introduce the slack variable $z \ge \mathop {\max }\limits_{u \in \mathcal{U} } \widetilde{P}^{u}_{\mathrm{out}}$, then the problem can be rewritten as
\begin{subequations}
\label{rp}
\begin{align}
  \mathop {\min}\limits_{\left\{t_c,\mathbf{t}\right\}} \quad & z \label{rp0} \\
 \text { s.t. } \;\;\; &   z \ge \widetilde{P}^{u}_{\mathrm{out}}, \forall u, \label{rp1} \\
&(\ref{p2}), (\ref{p3}).
\label{rp2}  
\end{align}        
\end{subequations}
Based on the (\ref{rp1}), we attempt to determine the structure and characteristics of the $\widetilde{P}^{u}_{\mathrm{out}}$, which leads to the following proposition.\\
\textbf{Proposition 4:} \textit{Let $\alpha>0$ and $D,M\in\mathbb{N}$. For each block $d\in\{1,\dots,D\}$ and node
$n\in\{1,\dots,M\}$, set $X_{d,n}\triangleq \sqrt{\kappa_d\,\xi_n}>0$, and let
$L_d\in\mathbb{N}$, $w_n^{(\alpha-1)}>0$, $\theta>0$, and $\rho_d<1$. Given any
$\gamma^u_{\mathrm{th}}>0$, we then define the following
\begin{subequations}
\label{eq:P2}
\begin{align}
Y_d(\gamma^u_{\mathrm{th}}) &\triangleq \sqrt{ \frac{2\,\gamma^u_{\mathrm{th}}}{\theta(1-\rho_d)} }, \\
S_d(\gamma^u_{\mathrm{th}}) &\triangleq \sum_{n=1}^{M} \Big[1-Q_\alpha\!\big(X_{d,n},Y_d(\gamma^u_{\mathrm{th}})\big)\Big]^{L_d}\, w_n^{(\alpha-1)}, \\
\widetilde{P}^{u}_{\mathrm{out}}(\gamma^u_{\mathrm{th}}) &\triangleq \prod_{d=1}^{D} \frac{1}{\Gamma(\alpha)}\, S_d(\gamma^u_{\mathrm{th}}).
\end{align}
\end{subequations}
Then, $\widetilde{P}^{u}_{\mathrm{out}}(\gamma^u_{\mathrm{th}})$ is strictly increasing in $\gamma^u_{\mathrm{th}}$ for all $\gamma^u_{\mathrm{th}}>0$.
}

\textit{Proof.} See Appendix C.

Based on the strictly increasing property proposed in \textbf{Proposition 4}, we can equivalently transfer the constraint (\ref{rp1}) into
\begin{equation}
  \gamma^u_{\mathrm{th}} \le \bar{\gamma}_u \left (z \right),  \forall u,
  \label{eq:inverse}
\end{equation}
where $\bar{\gamma}_u \left (z \right) \triangleq \left (P^{u}_{\mathrm{out}}\right)^{-1} \left (z \right) $ denotes as the per-user threshold requirement corresponding to the OP $z$. Combining the definition $\gamma^{u}_{\mathrm{th}}=\max\{\tilde{\gamma}^{c,u}_{\mathrm{th}},\,\tilde{\gamma}^{p,u}_{\mathrm{th}}\}$ with \eqref{eq:inverse} yields the following pair of feasibility constraints:
\begin{subequations}
\label{eq:newcons}
\begin{align}
t_u &\ge a_u \left (z\right), \label{newcons1} \\
t_c- \gamma^{c,u}_{\mathrm{th}} t_u & \ge b_u \left (z\right), \label{newcons2}
\end{align}
\end{subequations}
where $a_u \left (z\right)=\gamma^{p,u}_{\mathrm{th}}/\bar{\gamma } \bar{\gamma}_u \left (z \right)$ and $b_u \left (z\right) = \gamma^{c,u}_{\mathrm{th}}/\bar{\gamma } \bar{\gamma}_u \left (z \right)$, respectively. Hence, we can now rewrite the problem as
\begin{subequations}
\label{fp}
\begin{align}
  \mathop {\min}\limits_{\left\{t_c,\mathbf{t}\right\}}  z \label{fp0} \\
\text { s.t. }  \; \; &  (\ref{p2}), (\ref{p3}), (\ref{newcons1}), (\ref{newcons2}). \label{fp1}  
\end{align}        
\end{subequations}

To address the problem, we adopt a bisection search on the target level $z$. For any fixed $z$, we perform a feasibility check by seeking \((t_c,\mathbf{t})\) that satisfy all constraints. This leads to the following proposition:\\
\textbf{Proposition 5:}
Define the following linear program (LP)
\begin{subequations}
\label{lp}
\begin{align}
\mathop {\min}\limits_{t_{c}, \mathbf{t}} \quad & T=t_{c}+\sum_{u} t_{u} \\
 \text { s.t. } \;\;\; & t_{u} \geq a_{u}(z), \quad t_{c}- \gamma^{c,u}_{\mathrm{th}} t_{u} \geq b_{u}(z), \forall u, \\
& t_{c} \geq 0, t_{u} \geq 0
\end{align}        
\end{subequations}
Then, the level-$z$ feasibility condition holds \emph{if and only if} the optimal value $T^\star (z)$ of (\ref{lp}) satisfies $T^\star (z) \le 1$.
Moreover, whenever $T^\star (z) \le 1$, any optimizer $(t_c^\star,\mathbf{t}^\star)$ of \eqref{lp} constitutes a feasible allocation for level-$z$.

\textit{Proof.} If level-$z$ feasibility condition holds for some $(\hat t_c, \mathbf{\hat{t}_u})$, then $T^\star(z)\le \hat t_c+\sum_u \hat t_u\le 1$ by optimality of (\ref{lp}). Conversely, if $T^\star(z)\le 1$, any optimizer $(t_c^\star,\mathbf{t}_u^\star)$ of (\ref{lp}) satisfies the same linear constraints and $t_c^\star+\sum_u t_u^\star=T^\star(z)\le 1$, hence is feasible for level-$z$. 

Leveraging \textbf{Proposition~5}, solving (\ref{lp}) yields the optimizer $(t_c^\star,\mathbf{t}_u^\star)$ and the optimal value $T^\star(z)$, from which level-$z$ feasibility is certified by the criterion $T^\star(z)\le 1$.
The solution to the LP in \eqref{lp} is characterized by the following theorem.\\
\textbf{Theorem 2:} \textit{A closed-form solution to the LP in \eqref{lp} is given by
\begin{equation}\label{eq:opt}
t_u^\star(z)=a_u(z),\quad 
t_c^\star(z)=\max_{u}\big\{\gamma^{c,u}_{\mathrm{th}}a_u(z)+b_u(z)\,\big\},
\end{equation}
with optimal value
\begin{align}\label{eq:Tstar-opt}
T^\star(z)&=t_c^\star(z)+\sum_{u}{t}_u^\star(z)\notag \\
&=\max_{u}\{\,\gamma^{c,u}_{\mathrm{th}} a_u(z)+b_u(z)\}+\sum_{u}a_u(z).
\end{align}
}

\textit{Proof:} See Appendix D. 

The overall bisection with the closed-form optimal power allocation (BCOPA) algorithm that solves (\ref{rp}) is presented in \textbf{Algorithm 1}. Our proposed algorithm bisects a scalar outage level $z\in[0,1]$.  At each midpoint and for every user, the strictly increasing outage map is used to obtain $\bar{\gamma}_u(z)=\left(P^{u}_{\mathrm{out}}\right)^{-1} \left (z \right)$ through a 1-D search in $\gamma$. These inverses yield $a_u(z)=\gamma_{\mathrm{th}}^{p,u}/\big(\bar{\gamma}\,\bar{\gamma}_u(z)\big)$ and $b_u(z)=\gamma_{\mathrm{th}}^{c,u}/\big(\bar{\gamma}\,\bar{\gamma}_u(z)\big)$, from which the minimum power budget required at level $z$ is $T(z)=\max_{u}\!\left\{\gamma_{\mathrm{th}}^{c,u}\,a_u(z)+b_u(z)\right\}+\sum_{u} a_u(z).$ If $T(z)\le 1$, we have a feasible level and the upper bisection bound is updated. Otherwise, the lower bound is raised. Upon termination, the minimal feasible level $z^\star$ is obtained with logarithmic accuracy. The globally optimal power allocation follows the closed-form expressions of $t_u^\star=a_u(z^\star)$ for all $u$, and $t_c^\star=\max_{u}\!\left\{\gamma_{\mathrm{th}}^{c,u}\,a_u(z^\star)+b_u(z^\star)\right\}$. {By adaptively allocating power, $t_c^*$ guarantees the successful decoding of the common stream across all users in the network despite heterogeneous channel conditions.}


\begin{algorithm}[t]
\caption{Bisection with Closed-form Optimal Power Allocation}
\label{alg:bisection_outage}
\begin{algorithmic}[1]
\REQUIRE User set $\mathcal{U}$, per-user thresholds $\{\gamma_{\mathrm{th}}^{c,u},\gamma_{\mathrm{th}}^{p,u}\}_{u\in\mathcal{U}}$, SNR scale $\bar\gamma$, tolerances $\varepsilon_z, \varepsilon_\gamma$
\STATE Initialize bounds: $z_{\min}\gets 0$, $z_{\max}\gets 1$
\REPEAT
  \STATE $z \gets (z_{\min}+z_{\max})/2$
  \FORALL{$u\in\mathcal{U}$}

    \STATE $\bar\gamma_u(z)\gets (P^{u}_{\mathrm{out}})^{-1}(z)$, (solve $P^{u}_{\mathrm{out}}(\gamma)=z$ via bisection)

    \STATE $a_u(z)\gets \gamma_{\mathrm{th}}^{p,u}/\big(\bar\gamma\,\bar\gamma_u(z)\big)$,\quad $b_u(z)\gets \gamma_{\mathrm{th}}^{c,u}/\big(\bar\gamma\,\bar\gamma_u(z)\big)$
  \ENDFOR
  \STATE $t_c^{\min}(z)\gets \max_{u\in\mathcal{U}}\{\gamma_{\mathrm{th}}^{c,u}\,a_u(z)+b_u(z)\}$
  \STATE $T(z)\gets t_c^{\min}(z)+\sum_{u\in\mathcal{U}} a_u(z)$
  
  \STATE \textbf{if} $T(z)\le 1$ \textbf{then} $z_{\max}\gets z$ \textbf{else} $z_{\min}\gets z$
\UNTIL{$z_{\max}-z_{\min}\le \varepsilon_z$}

\STATE $z^\star\gets z_{\max}$, \quad $t_u^\star\gets a_u(z^\star)$, \quad $t_c^\star\gets \max_{u\in\mathcal{U}}\{\gamma_{\mathrm{th}}^{c,u}\,a_u(z^\star)+b_u(z^\star)\}$
\ENSURE Optimal variables $z^\star$, $t_c^\star$, $\{t_u^\star\}_{u\in\mathcal{U}}$
\end{algorithmic}
\end{algorithm}

\subsection{Convergence and complexity analysis}
\subsubsection{Convergence}
Let $z^\star:=\inf\{z\in[0,1]:T(z)\le1\}$ be the minimal feasible level. Starting from any $[z_{\min},z_{\max}]\subseteq[0,1]$ with $T(z_{\max})\le1$ and $T(z_{\min})>1$, the bisection iterates
$z_k=(z_{\min}^{(k)}\!+\!z_{\max}^{(k)})/2$
and update the feasible upper bound for $T(z_k)\le1$, otherwise the lower bound. Through standard bisection theory on monotone predicates, we have
\begin{equation}
z_{\max}^{(k)}-z_{\min}^{(k)} \le 2^{-k}\bigl(z_{\max}^{(0)}-z_{\min}^{(0)}\bigr),
z_{\min}^{(k)}\uparrow z^\star,\ z_{\max}^{(k)}\downarrow z^\star.
\end{equation}
Hence, the method converges \emph{linearly} with factor $1/2$ to $z^\star$. The stopping rule enforces $z_{\max}^{(k)}-z_{\min}^{(k)}\le\varepsilon_z$, then the returned $z_{\text{out}}=z_{\max}^{(k)}$ satisfies $0\le z_{\text{out}}-z^\star\le \varepsilon_z$.

\subsubsection{Complexity}
Let $U$ be the number of users and $D$ the number of correlation blocks. One outer iteration requires, for each user, an inner inversion of $\left (P^{u}_{\mathrm{out}}\right)^{-1} \left (z \right)$ by 1-D bisection:
\begin{equation}
N_{\text{inner}} = \Big\lceil \log\!\frac{1}{\varepsilon_\gamma}\Big\rceil.
\label{complexity}    
\end{equation}
Each evaluation of $\left (P^{u}_{\mathrm{out}}\right)^{-1} \left (z \right)$ factorizes over blocks:
\begin{itemize}
\item if $\rho_d\approx0$ (independent), the block cost is $O(1)$ using a Gamma CDF;
\item if $\rho_d\approx1$ (fully correlated), the block cost is $O(1)$ with a single Gamma CDF;
\item if $\rho_d\in(0,1)$, the block cost is $O(M_{\text{GL}})$ using Gauss-Laguerre polynomials.
\end{itemize}
Denote by $D_{\text{GL}}$ the number of partially correlated blocks (those with $\rho_d\in(0,1)$). Then, 1-D bisection evaluation costs $ O\big(D-D_{\text{GL}} + D_{\text{GL}} M_{\text{GL}}\big),$ and the overall arithmetic complexity is $O\big(\log\!\frac{1}{\varepsilon_z} \cdot U \cdot \log\frac{1}{\varepsilon_\gamma} \cdot \left ( D-D_{\text{GL}} + D_{\text{GL}} M_{\text{GL}} \right )\big)$.

\section{NUMERICAL RESULTS AND ANALYSIS}
In this section, we present a comprehensive numerical study across a range of system configurations and channel conditions. We first benchmark the OP of the proposed FAS-RSMA architecture against a receiver-side baseline, i.e., the FPA system. We then evaluate a fairness-driven min-max OP power-allocation policy and compare the results with the equal (non-optimized) power allocation case.

\subsection{System Parameters}
We normalize the per-antenna average power to $\Omega_u=1$. Unless otherwise specified, the default configuration employs a BS with $L=4$ transmit antennas serving $U=3$ users. {For the Monte Carlo simulations, correlated Nakagami-$m$ samples are generated according to the shape parameter $m$. When $m$ is an integer, we exploit the standard construction that a Nakagami-$m$ envelope can be obtained from the square root of the sum of $m$ independent Rayleigh envelopes. In particular, we first generate $m$ independent standard complex Gaussian vectors, impose the target spatial correlation through the Cholesky factorization of the target covariance matrix, and then form the channel envelopes by summing the squared magnitudes element-wise and taking the square root.
For non-integer $m$, we adopt an inverse-CDF approach based on a Gaussian copula, where the Gaussian correlation matrix is numerically calibrated in advance so that the resulting samples match the prescribed spatial correlation.} Each plotted point averages over $N_{\mathrm{sim}}=1 \times 10^{7}$ independent channel realizations.

\subsection{Performance of Outage Probability}
In this section, we validate the OP analysis of the FAS-RSMA system by comparing the closed-form predictions with Monte Carlo simulations under varying system parameters, target data-rate thresholds, and receiver-side antenna configurations. The power allocation for the three users is considered as fixed with $t_c=0.4$ and $t_1=t_2=t_3=0.2$. 

\begin{figure*}[ht]
\centering
\subfloat[Normalization parameter $W=2.5$\label{fig:1a}]{\includegraphics[width=0.4\linewidth]{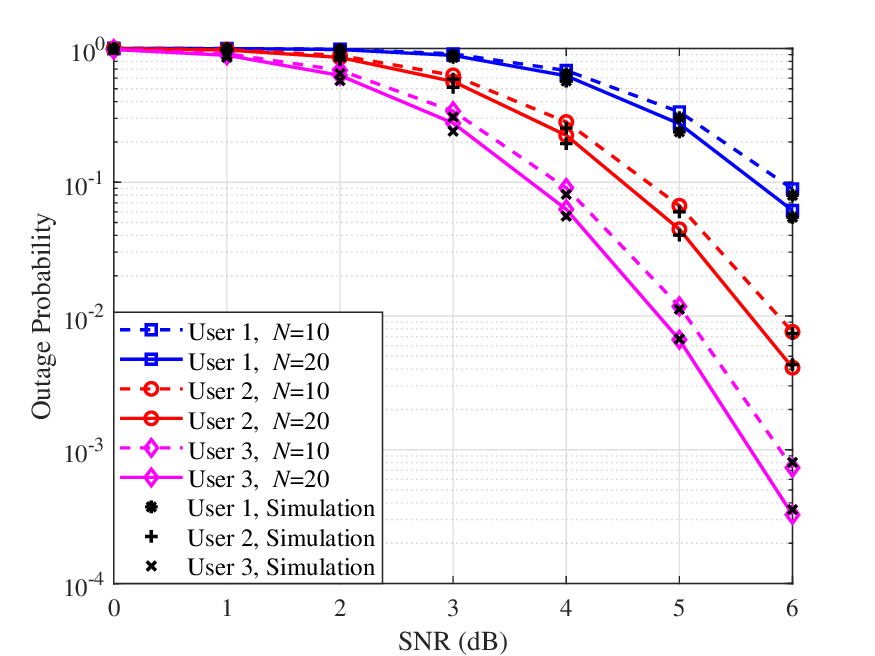}}
\hspace{0.02\linewidth}
\subfloat[Normalization parameter $W=4$\label{fig:1b}]{\includegraphics[width=0.4\linewidth]{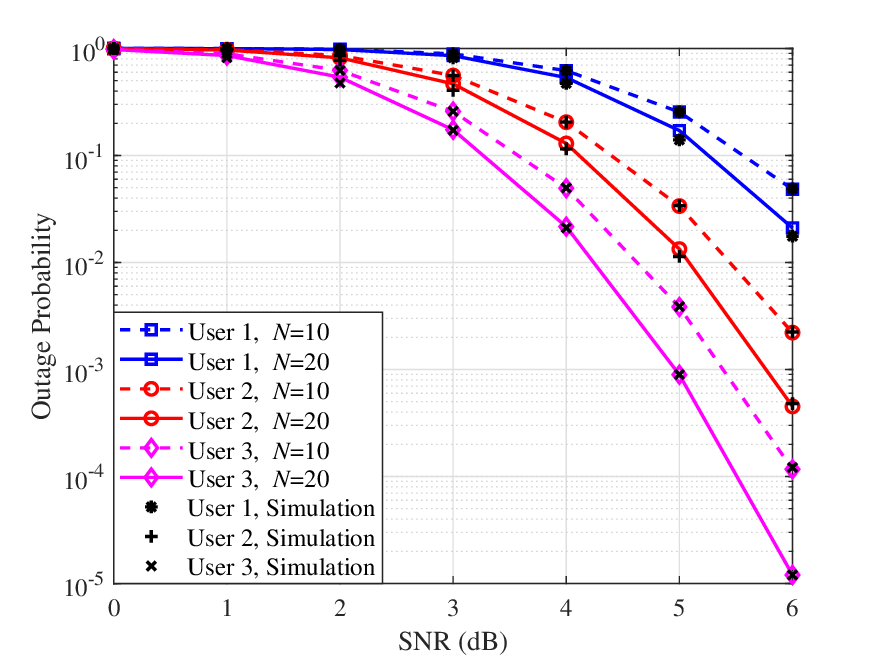}} 
\caption{OP versus SNR for three users under different FAS configurations with $\widetilde{R}_c=1$, $\widetilde{R}_1=2.3$, $\widetilde{R}_2=2.0$, and $\widetilde{R}_3=1.8$.}
\label{fig:1}
\end{figure*}

Fig.~\ref{fig:1} plots the OP versus SNR for three users under two normalization parameters \(W\in\{2.5,4\}\) and two FAS port counts \(N\in\{10,20\}\). Blue, red, and pink curves correspond to users 1, 2, and 3, respectively. The analytical curves closely follow the Monte Carlo markers in all cases, supporting the closed-form derivations. For a fixed \(W\), increasing the port number from \(N=10\) to \(N=20\) uniformly lowers the OP for every user and yields steeper high-SNR slopes. This trend indicates that a larger selection set can improve effective port-selection diversity, thereby enhancing the spatial diversity gain. For a fixed \(N\), raising \(W\) shifts all OP curves downward, indicating that a wider normalized aperture improves the chance of sampling favorable channel realizations (and effectively weakens spatial correlation). The user ordering (user~1 worst, user~3 best) is preserved across configurations due to their different private-rate targets, i.e., $\widetilde{R}_1=2.3$, $\widetilde{R}_2=2.0$, and $\widetilde{R}_3=1.8$. Overall, the results suggest that FAS-RSMA systems can improve OP by increasing the number of ports and adopting correlation-aware aperture layouts, thereby meeting reliability targets more efficiently under practical hardware constraints.

\begin{figure*}[ht]
\centering
\subfloat[Fading figure parameter $m=1$\label{fig:2a}]{\includegraphics[width=0.4\linewidth]{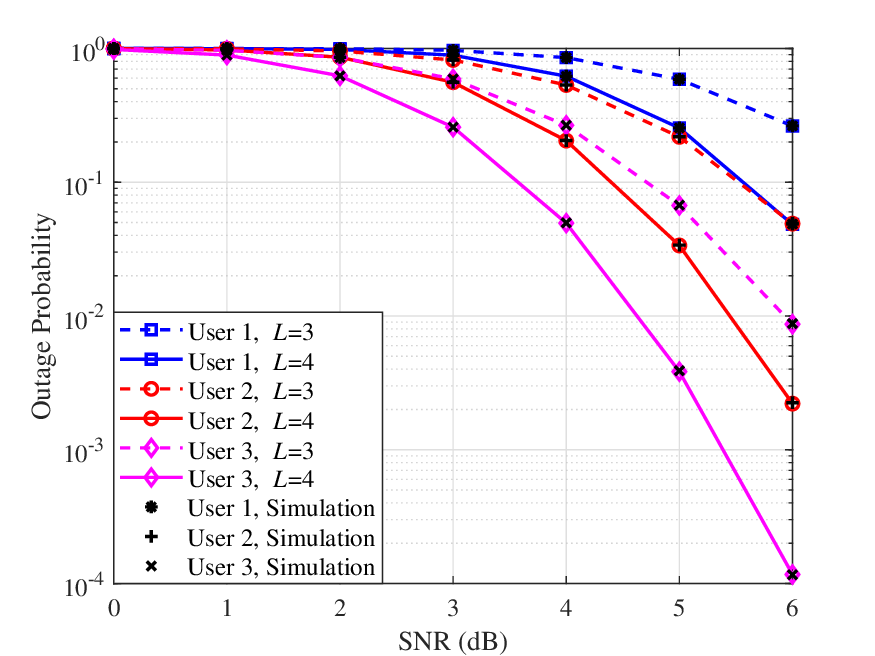}}
\hspace{0.02\linewidth}
\subfloat[Fading figure parameter $m=2$\label{fig:2b}]{\includegraphics[width=0.4\linewidth]{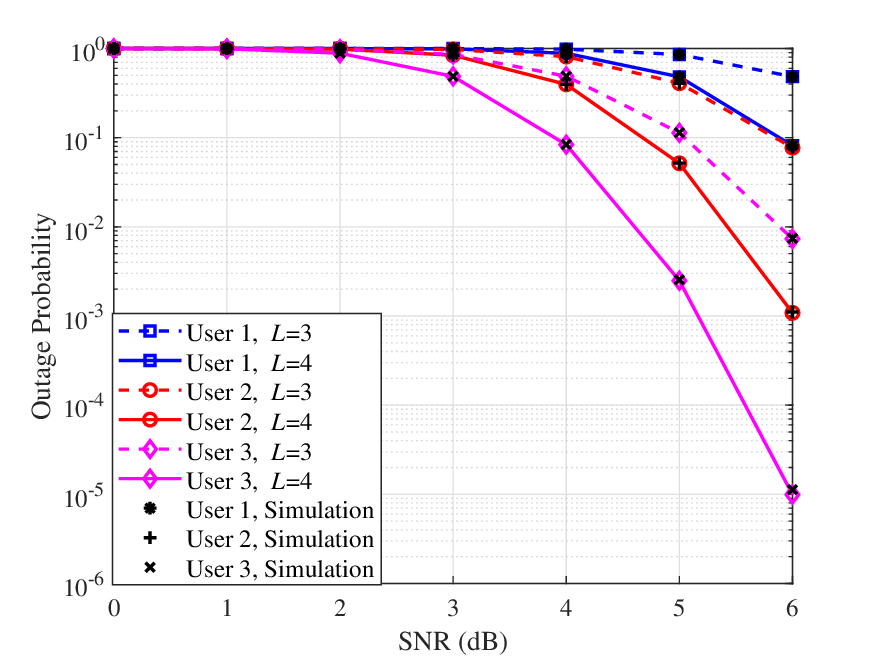}} 
\caption{OP versus SNR for three users under various channel conditions and antenna settings with $\widetilde{R}_c=1$, $\widetilde{R}_1=2.3$, $\widetilde{R}_2=2.0$, and $\widetilde{R}_3=1.8$.}
\label{fig:2}
\end{figure*}

Fig.~\ref{fig:2} reports per-user OP over Nakagami-$m$ channels for two fading severities ($m=1$ and $m=2$) and two transmit-antenna sizes ($L=3$ and $L=4$). The closed-form predictions continue to match with Monte Carlo simulation points, substantiating the analytical development. Increasing the array from $L=3$ to $L=4$ uniformly shifts the curves downward and steepens the high-SNR slope, signaling a higher effective diversity order due to the increased number of transmitted antennas. Moreover, the curves indicate that larger Nakagami parameters $m$ yield improved OP at high SNR. For a Nakagami-$m$ channel with fixed mean power \(\Omega_u=1\), increasing $m$ reduces the spread of the instantaneous power and lowers the probability of deep fades. From a physical standpoint, a higher $m$ corresponds to a stronger line-of-sight component and smaller amplitude fluctuations, which translates into better outage performance. Based on these results, we have the following practical insight: To meet stringent outage targets under practical link-budget constraints, additional transmit antennas and larger Nakagami parameters are both beneficial for improving reliability.

\begin{figure}[ht]
\setlength{\belowcaptionskip}{-0.9cm} %
\vskip 0.1in
\begin{center}
\centerline{\includegraphics[width=0.43\textwidth]{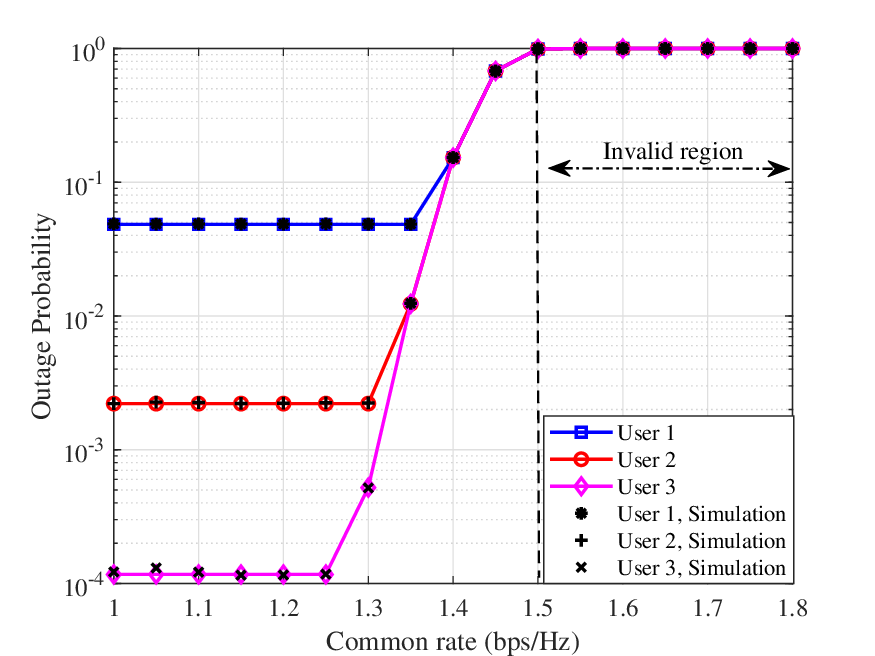}}
\caption{{OP comparison for $N=10$ ports under different common-rate thresholds with private-rate targets $\widetilde{R}_1=2.3$, $\widetilde{R}_2=2.0$, and $\widetilde{R}_3=1.8$.}}
\label{fig:3}
\end{center}
\vskip -0.1in
\end{figure}

The impact of the common and private data rate thresholds on OP performance is demonstrated in Fig.~\ref{fig:3}. For small common data rate thresholds, the OP of each user is essentially flat, being dominated by the private decoding condition. Accordingly, User~1 (most stringent private rate target with $\widetilde{R}_1=2.3$) exhibits the highest OP, whereas User~3 (most relaxed private rate target with $\widetilde{R}_3=1.8$) attains the lowest OP results. As the common data rate threshold increases, the OP of all the users exhibits a rise since the common stream becomes rate-limiting, where the common rate threshold cannot be satisfied based on the current power allocation strategy. Beyond the vertical demarcation, the \emph{invalid region} appears since the common data rate threshold violates the feasibility bound in Remark 1, the algebraic SINR becomes nonphysical, and outage occurs with probability one. These results indicate that the common-rate threshold should remain within the feasible region, with power reallocation if needed, to avoid reliability degradation. Before the common stream becomes rate-limiting, the user outage ordering is mainly governed by the private-rate targets and should thus reflect heterogeneous QoS requirements.

\begin{figure}[ht]
\setlength{\belowcaptionskip}{-0.9cm} %
\vskip 0.1in
\begin{center}
\centerline{\includegraphics[width=0.43\textwidth]{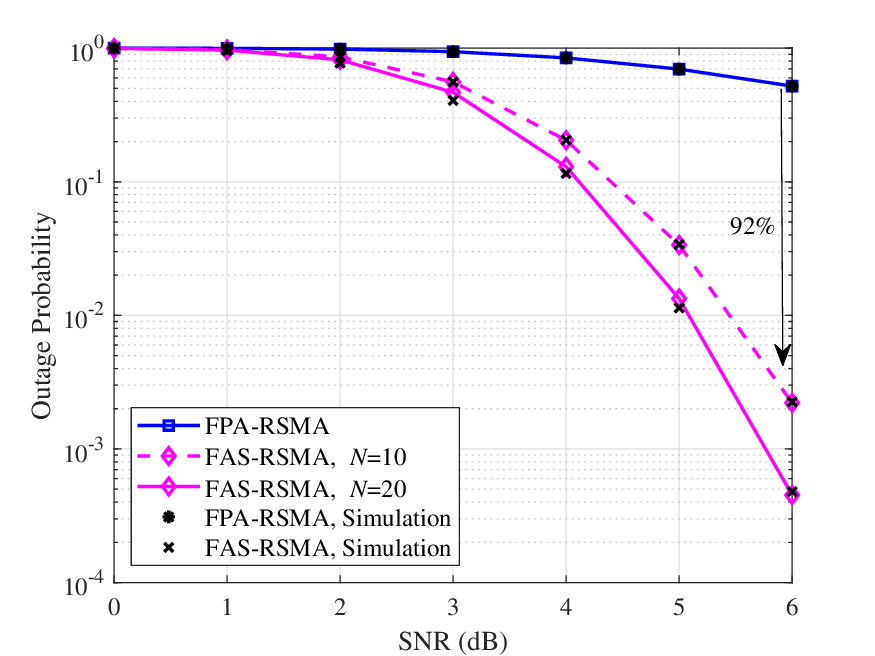}}
\caption{Performance of OP comparison between FAS and FPA of the RSMA system with $\widetilde{R}_c=1$, $\widetilde{R}_1=2.0$, $\widetilde{R}_2=2.0$, and $\widetilde{R}_3=2.0$.}
\label{fig:4}
\end{center}
\vskip -0.1in
\end{figure}

Fig.~\ref{fig:4} reports OP versus SNR for different RSMA receiver architectures, i.e., FPA baseline with a single, non-reconfigurable antenna per user, and the proposed FAS-RSMA with $N\!\in\!\{10,20\}$ antenna ports. Under identical power splits and identical common and private data rate thresholds across users, all users exhibit the same OP, so each architecture appears as a single curve. Across the evaluated SNR range, the FAS-RSMA curve (magenta) lies strictly below the benchmark, delivering up to $\approx 92\%$ lower OP than FPA-RSMA (blue). Compared with FPA-RSMA, the performance gain of FAS-RSMA stems from a larger pool of candidate ports. Although closely spaced antenna ports introduce correlation for FAS, the increased port number yields port-selection diversity. The FPA baseline performs poorly since it provides neither selection nor reconfiguration. Moreover, the gap between FAS-RSMA and FPA-RSMA widens as the number of ports $N$ increases, indicating that FAS scalability can be effectively exploited to further improve system reliability. These results indicate that, despite the spatial correlation induced by closely spaced ports, FAS can still realize \emph{additional port-selection diversity}, which in turn improves reliability compared with the FPA system.

\subsection{Optimization of Min-Max Outage Probability}
In this section, we present simulation results for the proposed BCOPA algorithm applied to the \emph{min-max} OP problem, yielding the fairness common and private power allocation. We benchmark BCOPA against (i) fixed, non-optimized power splits and (ii) representative multiple-access baselines.

\begin{figure}[ht]
\setlength{\belowcaptionskip}{-0.9cm} %
\vskip 0.1in
\begin{center}
\centerline{\includegraphics[width=0.43\textwidth]{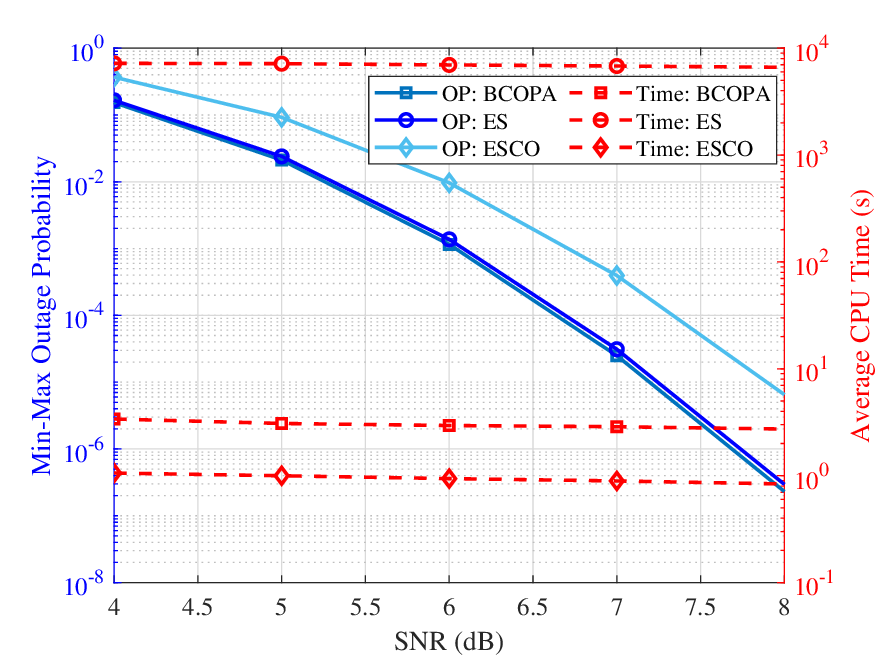}}
\caption{{Performance comparison of min-max OP and computational complexity among BCOPA, ES, and ESCO for the FAS-RSMA system.} }
\label{fig:bcopa}
\end{center}
\vskip -0.1in
\end{figure}

{Fig.~\ref{fig:bcopa} contrasts the min-max OP (left y-axis) and the average CPU time (right y-axis) of the proposed BCOPA algorithm against the ES algorithm and a simplified exhaustive-search baseline with common-only optimization (ESCO), where private streams share equal power and only the common-stream power is optimized. As observed, the min-max OP curves of BCOPA and ES overlap across the entire SNR range, indicating that BCOPA attains the globally optimal solution. In contrast, the ESCO baseline exhibits a noticeably degraded min-max OP (cyan curve). This performance gap arises because the fairness problem considered herein is not fully symmetric, whose heterogeneous private-rate requirements ($\tilde{R}_{p,u}$) induce different outage thresholds. Forcing equal private power allocation creates a bottleneck at the user with the most stringent data requirement. Regarding computational cost, BCOPA completes in $\mathcal{O}(10^{0})$ s per instance. While ESCO is marginally faster due to its reduced search space, its performance degradation makes it sub-optimal for the fairness objective. Meanwhile, the full ES requires about $\mathcal{O}(10^{4})$ s. These results show that our proposed BCOPA preserves optimality while drastically reducing the runtime, dynamically adapting both common and private powers to diverse user requirements.}

\begin{figure}[ht]
\setlength{\belowcaptionskip}{-0.9cm} %
\vskip 0.1in
\begin{center}
\centerline{\includegraphics[width=0.43\textwidth]{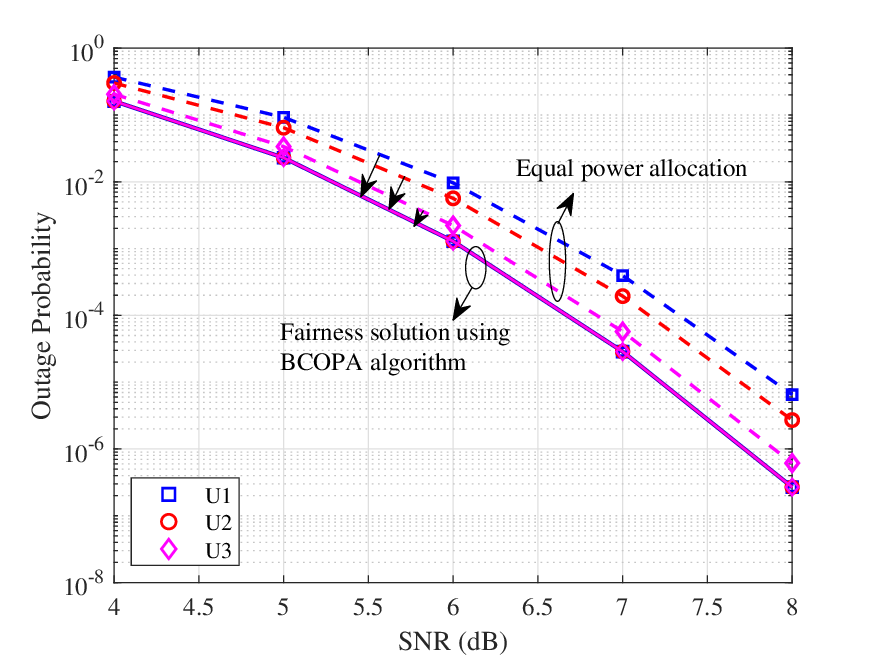}}
\caption{Performance of per-user OP between BCOPA-designed method and equal power allocation method with $\widetilde{R}_c=1$, $\widetilde{R}_1=2.3$, $\widetilde{R}_2=2.2$, and $\widetilde{R}_3=1.5$. }
\label{fig:bcopa_fair}
\end{center}
\vskip -0.1in
\end{figure}

Fig.~\ref{fig:bcopa_fair} demonstrates the per-user OP for three users under
baseline equal power splits (dashed line) where $t_c=0.3$ and $t_1=t_2=t_3= (1-t_c) /3$, and the proposed BCOPA-designed common and private power allocation (solid line). Without optimization, a clear disparity is observed across users. User~1 consistently exhibits the highest OP, whereas User~3 achieves the lowest, and the gap persists over the entire SNR range. In contrast, BCOPA drives the system to the min-max OP operating point: the three solid curves coincide at each SNR, indicating that all users attain the \emph{same} OP, and the worst user improved by  $1$ dB SNR gain at a fixed OP, thereby guaranteeing the user fairness. Moreover, the BCOPA curve lies strictly below every baseline curve for all SNRs, showing that the optimized allocation simultaneously improves the reliability of the users rather than merely equalizing their performance. 
This gain comes from the coupled common and private streams outage constraints, under which power reallocation relaxes one of the decoding conditions and reduces the OP of all users. Hence, the proposed BCOPA algorithm provides a fairness-optimal and practical solution for QoS provisioning in FAS-RSMA systems.

\begin{figure}[ht]
\setlength{\belowcaptionskip}{-0.9cm} %
\vskip 0.1in
\begin{center}
\centerline{\includegraphics[width=0.43\textwidth]{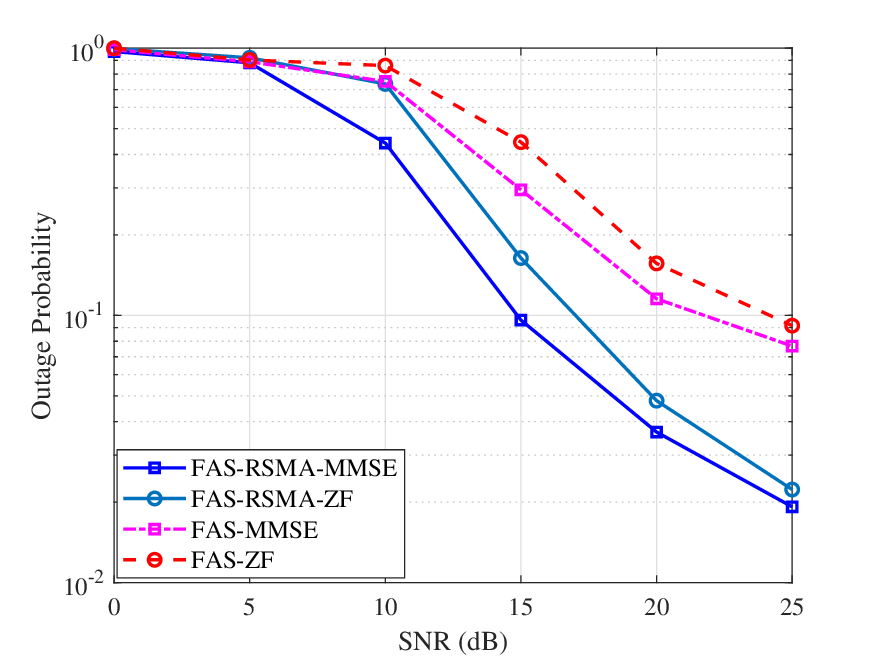}}
\caption{{OP comparison among the proposed FAS-RSMA scheme and conventional FAS-ZF and FAS-MMSE precoding schemes under $L=32$ and $U=16$.} }
\label{fig:comp}
\end{center}
\vskip -0.1in
\end{figure}

{To further demonstrate the advantage of the proposed FAS-RSMA over conventional SDMA with linear precoding under a more practical BS array size, we consider an additional scenario with $L=32$ BS antennas serving $U=16$ FAS users. The users are divided into two groups: users within the same group have highly aligned channel directions, whereas the two groups remain distinguishable. Imperfect CSIT is also taken into account \cite{ref7}, such that the precoders are designed based on the estimated channels and their performance is evaluated on the corresponding true channels. For a fair comparison, all schemes employ the same FAS configuration and total transmit-power budget. In addition to the conventional FAS-ZF and FAS-MMSE benchmarks, we also evaluate two RSMA realizations, namely FAS-RSMA-ZF and FAS-RSMA-MMSE, where the private streams are designed by ZF-type and MMSE-type precoders, respectively. As shown in Fig.~\ref{fig:comp}, both FAS-RSMA variants achieve lower OP than the corresponding conventional baselines, while FAS-RSMA-MMSE provides the best performance overall. This performance gap arises since conventional linear precoders are sensitive to residual interference when user channels are aligned and CSIT is imperfect, whereas RSMA can partially shift interference-limited transmission to the common stream. This confirms that RSMA offers a clear robustness advantage when the user channels are spatially clustered and the CSIT is imperfect.}

\section{Conclusion}
This work developed an analytical framework for multiuser MISO downlink FAS-RSMA systems with spatially correlated ports. Closed-form OP expressions were derived based on a block-correlation model, hybrid ZF-MRT precoding, and generalized Gauss-Laguerre quadrature. A min-max OP design was then formulated and solved by the proposed BCOPA algorithm. Numerical results validated the analysis, demonstrated the OP advantage of FAS-RSMA over the FPA baseline, and showed that the fairness design improves worst-user reliability. Future work will extend the framework to two-dimensional apertures, imperfect CSI and hardware impairments, and joint transceiver optimization.

\appendices
\section{Proof of Proposition 2}
For a single complex Gaussian $X\sim\mathcal{CN}(\mu,\sigma^{2})$, the real-imaginary decomposition implies
\begin{equation}
\frac{2}{\sigma^{2}}\lvert X\rvert^{2}
\;\sim\; \chi_{2}^{\prime\,2}\!\Big(\lambda=\tfrac{2}{\sigma^{2}}\lvert\mu\rvert^{2}\Big).
\end{equation}
For $\{X_q\}_{q=1}^{\alpha}$ independent with $X_q\sim\mathcal{CN}(\mu_q,\sigma^{2})$, additivity of noncentral chi-square variables with common scale yields
\begin{equation}
\frac{2}{\sigma^{2}}\sum_{q=1}^{\alpha}\lvert X_q\rvert^{2}
\;\sim\; \chi_{2\alpha}^{\prime\,2}\!\Big(\lambda=\tfrac{2}{\sigma^{2}}\sum_{q=1}^{\alpha}\lvert\mu_q\rvert^{2}\Big).
\end{equation}
Based on the (\ref{eq:aa}), we can have the following definition for the sum of $\lvert\mu_q\rvert^{2}$ as
\begin{equation}
\sum_{q=1}^{\alpha}\lvert\mu_q\rvert^{2}
=\theta\rho_d\sum_{q=1}^{\alpha}\lvert z_{0,q}\rvert^{2}
=\theta\rho_d\,u,
\end{equation}
Applying this with $X_q=\hat{h}^{\,d}_{i,q}$ and $\sigma^{2}=\sigma_d^{2}$ gives the following conditional on $U_d=u$:
\begin{equation}
\frac{2}{\sigma_d^{2}}\sum_{q=1}^{\alpha}\bigl|\hat{h}^{\,d}_{i,q}\bigr|^{2}
\;\sim\;
\chi_{2\alpha}^{\prime\,2}\!\Big(\lambda=\tfrac{2\,\theta\rho_d}{\sigma_d^{2}}\,u\Big).
\end{equation}
Then, by using $\sigma_d^{2}=\theta(1-\rho_d)$, the noncentrality simplifies to
\begin{equation}
\lambda \;=\; \frac{2\,\theta\rho_d}{\theta(1-\rho_d)}\,u \;=\; \frac{2\rho_d}{1-\rho_d}\,u \;=\; \kappa_d\,u.
\end{equation}
Therefore, we have
\begin{equation}
X^{d}_{i}\,\big|\,U_d=u
=\sum_{q=1}^{\alpha}\bigl|\hat{h}^{\,d}_{i,q}\bigr|^{2}
\ \stackrel{d}{=}\ 
\frac{\sigma_d^{2}}{2}\;\chi_{2\alpha}^{\prime\,2}\!\big(\lambda=\kappa_d u\big),
\end{equation}
which is exactly \eqref{eq:cond-ncx2}. This completes the proof.

\section{Proof of Proposition 3}

{Suppose that $Y\sim\chi_{2M}^{\prime 2}(\lambda)$. For $M>0$, its CDF is given by
\begin{equation}
\Pr\{Y\le y\}=1-Q_M\!\big(\sqrt{\lambda},\sqrt{y}\big),
\end{equation}
where $Q_M(\cdot,\cdot)$ is the generalized Marcum $Q$-function~\cite{ref48}. According to \textbf{Proposition~2}, the conditional CDF of $X_i^d$ conditioned on $U_d=v$ can be written as
\begin{equation}
F_{X_i^d|U_d=v}(x)
=
\Pr\{X_i^d\le x\mid U_d=v\}
=
1-Q_{\alpha}\!\Big(\sqrt{\kappa_d v},\sqrt{\tfrac{2x}{\sigma_d^2}}\Big).
\label{eq:cond-cdf-app}
\end{equation}
Since the $L_d$ ports in block $d$ are conditionally i.i.d. given $U_d=v$, the conditional CDF of the blockwise maximum
$X^d\triangleq \max_{1\le i\le L_d} X_i^d$
is
\begin{equation}
F_{X^d|U_d=v}(x)
=
\Big[F_{X_i^d|U_d=v}(x)\Big]^{L_d}.
\label{eq:block-max-cond-app}
\end{equation}
By averaging over the common Gamma random variable $U_d\sim\Gamma(\alpha,1)$, the unconditional CDF of $X^d$ is obtained as
\begin{align}
F_{X^d}(x)
&=
\int_{0}^{\infty} F_{X^d|U_d=v}(x) f_{U_d}(v)\,dv
\notag\\
&=
\int_{0}^{\infty}
\Big[1-Q_{\alpha}\!\Big(\sqrt{\kappa_d v},\sqrt{\tfrac{2x}{\theta(1-\rho_d)}}\Big)\Big]^{L_d}
\frac{v^{\alpha-1}e^{-v}}{\Gamma(\alpha)}\,dv.
\label{eq:block-cdf-app}
\end{align}
Finally, under the independence assumption across different block-correlation blocks, the CDF of the selected-port channel gain is given by
\begin{equation}
F_{|\zeta_u|^2}(x)=\prod_{d=1}^{D} F_{X^d}(x).
\label{eq:sel-cdf-app}
\end{equation}
Substituting $x=\gamma_{\mathrm{th}}^{u}$ into \eqref{eq:sel-cdf-app} directly yields the OP expression in \eqref{eq:OP-final}. This completes the proof.}

\section{Proof of Proposition 4}
Based on the definition in (\ref{eq:P2}), we can take the derivative of the $P^{u}_{\mathrm{out}}$ with respect to $\gamma_{\mathrm{th}}^{u}$, which gives the following expression:
\begin{align}
\frac{d}{d\gamma_{\mathrm{th}}^{u}}P^{u}_{\mathrm{out}}(\gamma_{\mathrm{th}}^{u})
&=\Big(\prod_{d=1}^{D}\tfrac{1}{\Gamma(\alpha)}\Big)\!
\sum_{i=1}^{D}\Big(\prod_{d\neq i} S_d(\gamma_{\mathrm{th}}^{u})\Big)\, S_i'(\gamma_{\mathrm{th}}^{u})
 \notag\\
&= P^{u}_{\mathrm{out}}(\gamma_{\mathrm{th}}^{u})\;\sum_{d=1}^{D}\frac{S_d'(\gamma_{\mathrm{th}}^{u})}{S_d(\gamma_{\mathrm{th}}^{u})},
\label{eq:Fprime-log}
\end{align}
where it can be found that $P^{u}_{\mathrm{out}} >0$ and $S_d >0$. Hence, it is essential to determine the sign of the $S_d'(\gamma_{\mathrm{th}}^{u})$. We define $g_{d,n}(\gamma)\triangleq 1-Q_\alpha\!\big(X_{d,n},Y_d(\gamma)\big)$, then the following expressions can be given
\begin{align}
& S_d'(\gamma_{\mathrm{th}}^{u})
\!\!=\!\!\sum_{n=1}^{M} L_d\,\big[g_{d,n}(\gamma_{\mathrm{th}}^{u})\big]^{L_d-1}\!\!
\frac{d}{d\gamma_{\mathrm{th}}^{u}}g_{d,n}(\gamma_{\mathrm{th}}^{u})\; w_n^{(\alpha-1)},
\label{eq:Sdprime-raw-en}\\
&\frac{d}{d\gamma_{\mathrm{th}}^{u}}g_{d,n}(\gamma_{\mathrm{th}}^{u})
= -\,\frac{\partial Q_\alpha}{\partial Y}\Big(X_{d,n},Y_d(\gamma_{\mathrm{th}}^{u})\Big)
\cdot \frac{dY_d}{d\gamma_{\mathrm{th}}^{u}} .
\label{eq:gprime-chain-en}
\end{align}

To determine the sign of the $S_d'(\gamma_{\mathrm{th}}^{u})$, We use two basic facts about the (generalized) Marcum--$Q$ function $Q_\alpha(X,Y)$ with $X>0$, $Y>0$, and $\alpha>0$:
\begin{itemize}
\item[(i)] $Q_\alpha(X,Y)$ is continuously differentiable in $Y>0$.
\item[(ii)] Its derivative of $Y$ has the following expression, which is based on Leibniz’s rule:
\begin{equation}
\frac{\partial}{\partial Y} Q_\alpha(X,Y)
= -\, Y\Big(\tfrac{Y}{X}\Big)^{\alpha-1} \exp\!\Big(-\tfrac{Y^2+X^2}{2}\Big) I_{\alpha-1}(XY) \label{eq:dQdY}
\end{equation}
where $I_{\alpha-1}(\cdot)$ is the modified Bessel function. Since $I_{\alpha-1}(z)>0$ for $z>0$, the right-hand side of \eqref{eq:dQdY} is strictly negative for all $X>0$, $Y>0$, $\alpha>0$. Hence, $Q_\alpha(X,Y)$ is strictly decreasing in $Y$.
\end{itemize}
Based on the two facts and $Y_d'(\gamma^u_{\mathrm{th}})=\frac{1}{2}\sqrt{\frac{2}{\theta(1-\rho_d)}}\,\left({\gamma^u_{\mathrm{th}}}\right) ^{-1/2}>0$, we can then determine $\frac{d}{d\gamma_{\mathrm{th}}^{u}}g_{d,n}(\gamma_{\mathrm{th}}^{u})>0$. Next, combining with the $w_n^{(\alpha-1)}>0$ and $L_d\big[g_{d,n}(\gamma_{\mathrm{th}}^{u})\big]^{L_d-1}>0$ imply that the $S_d'(\gamma_{\mathrm{th}}^{u}) >0$. Hence, $P^{u}_{\mathrm{out}}$ is strictly increasing in $\gamma^u_{\mathrm{th}}>0$. At $\gamma^u_{\mathrm{th}}=0$, $P^{u}_{\mathrm{out}}$ is continuous and the right derivative is nonnegative, hence $P^{u}_{\mathrm{out}}$ is nondecreasing at $\gamma^u_{\mathrm{th}}=0$, which proves the stated result.

\section{Proof of Theorem 2}
Problem~\eqref{lp} is a linear program and therefore convex. A strictly feasible point exists: choose any $t_u>a_u(z)$ for all $u$ and set $t_c>\max_{u}\{\gamma^{c,u}_{\mathrm{th}} t_u+b_u(z)\}$; all inequalities then hold with strict slack. Hence, Slater’s condition holds, strong duality applies, and the KKT conditions are necessary and sufficient for optimality.

We can write the Lagrangian with multipliers $\lambda_u,\mu_u\ge 0$ as
\begin{align}\label{eq:Lagr}
\mathcal{L}(t_c,\{t_u\},\lambda,\mu)
&= t_c+\sum_u t_u
+\sum_u \lambda_u\big(b_u(z)+\gamma^{c,u}_{\mathrm{th}} t_u-t_c\big) \notag\\
&+\sum_u \mu_u\big(a_u(z)-t_u\big).
\end{align}
Hence, we can have the following KKT conditions:
\begin{subequations}
\begin{align}
&t_u\ge a_u(z),\quad t_c\ge \gamma^{c,u}_{\mathrm{th}} t_u+b_u(z),\ \forall u; \label{eq:pf}\\
&\lambda_u\ge 0,\ \mu_u\ge 0,\ \forall u; \label{eq:df}\\
&\frac{\partial\mathcal{L}}{\partial t_c}=1-\sum_u \lambda_u=0,\quad
\frac{\partial\mathcal{L}}{\partial t_u}=1+\gamma^{c,u}_{\mathrm{th}}\lambda_u-\mu_u=0,\ \forall u; \label{eq:stat}\\
&
\lambda_u\big(b_u(z)+\gamma^{c,u}_{\mathrm{th}} t_u-t_c\big)=0,\quad
\mu_u\big(a_u(z)-t_u\big)=0,\ \forall u. \label{eq:cs}
\end{align}    
\end{subequations}

From~\eqref{eq:stat}, $\mu_u=1+\gamma^{c,u}_{\mathrm{th}}\lambda_u\ge 1$, hence $\mu_u>0$ for all $u$. Via~\eqref{eq:cs}, $a_u(z)-t_u=0$ for all $u$ gives
\begin{equation}\label{eq:tu-star}
t_u^\star=a_u(z),\forall u.
\end{equation}
With~\eqref{eq:tu-star}, primal feasibility~\eqref{eq:pf} reduces to $t_c\ge \gamma^{c,u}_{\mathrm{th}} a_u(z)+b_u(z)$ for all $u$. We define the following set
\begin{equation}\label{eq:I}
\mathcal I \triangleq \arg\max_{u}\{\,\gamma^{c,u}_{\mathrm{th}} a_u(z)+b_u(z)\,\}.
\end{equation}
Complementary slackness in~\eqref{eq:cs} implies $\lambda_u>0$ only for $u\in\mathcal I$ and $\lambda_u=0$ for $u\notin\mathcal I$. Stationarity in $t_c$ requires $\sum_u\lambda_u=1$, so we choose any probability vector $(\lambda_u)_{u\in\mathcal I}$ supported on $\mathcal I$. The only value of $t_c$ compatible with tightness on $\mathcal I$ and slack elsewhere is
\begin{equation}\label{eq:opt1}
t_c^\star=\max_{u}\{\,\gamma^{c,u}_{\mathrm{th}} a_u(z)+b_u(z)\,\},
\end{equation}
which yields~\eqref{eq:opt}.

\medskip

Evaluating the primal objective at $(t_c^\star,\{t_u^\star\})$ gives~\eqref{eq:Tstar-opt}. Alternatively, minimizing~\eqref{eq:Lagr} over $(t_c,\{t_u\})$ using~\eqref{eq:stat} yields the dual objective
\begin{align}\label{eq:dual}
g(\lambda,\mu)&=\sum_u \mu_u a_u(z)+\sum_u \lambda_u b_u(z) \notag \\
&=\sum_u a_u(z)+\sum_u \lambda_u\big(\gamma^{c,u}_{\mathrm{th}} a_u(z)+b_u(z)\big)  \notag\\
&=\sum_u a_u(z)+\max_{u}\{\gamma^{c,u}_{\mathrm{th}} a_u(z)+b_u(z)\},
\end{align}
where the last equality uses that $(\lambda_u)$ is a probability vector supported on $\mathcal I$. The primal and dual values coincide, confirming optimality.


 




\vfill


\begin{thebibliography}{1}
\bibliographystyle{IEEEtran}

\bibitem{ref1}
W. K. New \emph{et al.}, ``A tutorial on fluid antenna system for 6G networks: Encompassing communication theory, optimization methods and hardware designs," {\it{IEEE Commun. Surv. Tutor.}}, vol. 27, no. 4, pp. 2325-2377, Aug. 2025

\bibitem{ref2}
T. Wu \emph{et al.}, “Fluid antenna systems enabling 6G: Principles, applications, and research directions,” {\it{IEEE Wireless Commun.}}, early access, 2025, doi: 10.1109/MWC.2025.3629597.

\bibitem{ref3}
Y. Mao, O. Dizdar, B. Clerckx, R. Schober, P. Popovski, and H. V. Poor, ``Rate-splitting multiple access: Fundamentals, survey, and future research trends," {\it{IEEE Commun. Surv. Tutor.}}, vol. 24, no. 4, pp. 2073-2126, 2022.

\bibitem{ref4}
T. Han and K. Kobayashi, ``A new achievable rate region for the interference channel," {\it{IEEE Trans. Inf. Theory}}, vol. 27, no. 1, pp. 49-60, 1981.

\bibitem{ref5}
B. Clerckx, H. Joudeh, C. Hao, M. Dai, and B. Rassouli, ``Rate splitting for MIMO wireless networks: A promising PHY-layer strategy for LTE evolution,"  {\it{IEEE Commun. Mag.}}, vol. 54, no. 5, pp. 98-105, 2016.

\bibitem{ref6}
Y. Mao, B. Clerckx, and V. O. K. Li, “Rate-splitting multiple access for downlink communication systems: Bridging, generalizing, and outperforming SDMA and NOMA,” {\it{EURASIP J. Wireless Commun. Netw.}}, vol. 2018, no. 1, p. 133, May. 2018.

\bibitem{ref7}
H. Joudeh and B. Clerckx, “Sum-rate maximization for linearly precoded downlink multiuser MISO systems with partial CSIT: A rate-splitting approach,” {\it{IEEE Trans. on Comm.}}, vol. 64, no. 11, pp. 4847-4861, Nov. 2016.

\bibitem{ref8}
M. Dai, B. Clerckx, D. Gesbert, and G. Caire, “A rate splitting strategy for massive MIMO with imperfect CSIT,” {\it{IEEE Trans. Wireless Commun.}}, vol. 15, no. 7, pp. 4611-4624, Jul. 2016.

\bibitem{ref9}
Y. Mao, B. Clerckx, J. Zhang, V. O. Li, and M. A. Arafah, “Max-min fairness of K-user cooperative rate-splitting in MISO broadcast channel with user relaying,” {\it{IEEE Trans. Wireless Commun.}}, vol. 19, no. 10,
pp. 6362-6376, Oct. 2020.

\bibitem{ref10}
B. Clerckx, Y. Mao, R. Schober, and H. V. Poor, “Rate-splitting unifying SDMA, OMA, NOMA, and multicasting in MISO broadcast channel: A simple two-user rate analysis,” {\it{IEEE Wireless Commun. Lett.}}, vol. 9, no. 3, pp. 349-353, Nov. 2019.

\bibitem{ref11}
A. Bansal \emph{et al.}, ``Rate-splitting multiple access for intelligent reflecting surface aided multi-user communications," {\it{IEEE Trans. Veh. Technol.}}, vol. 70, no. 9, pp. 9217-9229, Sep. 2021.

\bibitem{ref12}
A. Mishra, Y. Mao, O. Dizdar, and B. Clerckx, “Rate-splitting multiple access for downlink multiuser MIMO: Precoder optimization and PHY layer design,” {\it{IEEE Trans. Commun.}}, vol. 70, no. 2, pp. 874-890, Feb. 2022.

\bibitem{ref13}
B. Clerckx \emph{et al.}, ``A primer on rate-splitting multiple access: Tutorial, myths, and frequently asked questions," {\it{IEEE J. Sel. Areas Commun.}}, vol. 41, no. 5, pp. 1265-1308, 2023.


\bibitem{ref15}
S. Zhang, B. Clerckx, D. Vargas, O. Haffenden, and A. Murphy, “Rate-splitting multiple access: Finite constellations, receiver design, and SIC-free implementation,” {\it{IEEE Trans. Commun.}}, vol. 72, no. 9, pp. 5319-5333, Sep. 2024

\bibitem{ref16}
J. Liu \emph{et al.}, ``Power allocation for high mobility OTFS-RSMA system with path-selective one-tap receiver," {\it{IEEE Wireless Commun. Lett.}}, vol. 14, no. 3, pp. 661-665, Mar. 2025.

\bibitem{ref17}
J. Park, B. Lee, J. Choi, H. Lee, N. Lee et al., ``Rate-splitting multiple access for 6G networks: Ten promising scenarios and applications,” {\it{IEEE Network}}, vol. 38, no. 3, pp. 128-136, Oct. 2023.

\bibitem{ref185}
T. -H. Vu \emph{et al.},``Outage, capacity, and error performance of downlink RSMA-based systems: Analysis and resource optimization," {\it{IEEE Trans. Commun.}}, vol. 73, no. 8, pp. 6868-6883, Aug. 2025.


\bibitem{ref20}
S. Zhang \emph{et al.}, ``FARS: Elevating rate-splitting multiple access in non-territorial networks with intelligent fluid antenna system," {\it{IEEE J. Sel. Areas Commun.}}, vol. 44, pp. 1045-1060, Sep. 2025. 

\bibitem{ref21}
K. K. Wong, A. Shojaeifard, K.-F. Tong, and Y. Zhang, ``Performance limits of fluid antenna systems," {\it{IEEE Commun. Lett.}}, vol. 24, no. 11, pp. 2469-2472, Nov. 2020.

\bibitem{ref22}
K.-K. Wong \emph{et al.}, ``Fluid antenna system," {\it{IEEE Trans. Wireless Commun.}}, vol. 20, no. 3, pp. 1950-1962, Mar 2021.

\bibitem{ref23}
P. Ram´ırez-Espinosa, D. Morales-Jimenez, and K. K. Wong, ``A new spatial block-correlation model for fluid antenna systems," {\it{IEEE Trans. Wireless Commun.}}, vol. 23, no. 11, pp. 15829-15843, Nov. 2024.

\bibitem{ref24}
B. Liu, K.-F. Tong, K. K. Wong, C.-B. Chae, and H. Wong, ``Programmable meta-fluid antenna for spatial multiplexing in fast fluctuating radio channels," {\it{Opt. Express}}, vol. 33, no. 13, pp. 28898-28915, 2025.

\bibitem{ref25}
Y. Shen \emph{et al.}, ``Design and implementation of mmWave surface wave enabled fluid antennas and experimental results for fluid antenna multiple access," {\it{arXiv preprint}}, arXiv:2405.09663, May 2024.

\bibitem{ref26}
J. Zhang \emph{et al.}, ``A novel pixel-based reconfigurable antenna applied in fluid antenna systems with high switching speed," {\it{IEEE Open J. Antennas Propag.}}, vol. 6, no. 1, pp. 212-228, Feb. 2025.

\bibitem{ref27}
K.-K. Wong and K.-F. Tong, ``Fluid antenna multiple access," {\it{IEEE Trans. Wireless Commun.}}, vol. 21, no. 7, pp. 4801-4815, Jul. 2022.


\bibitem{ref29}
F. R. Ghadi \emph{et al.}, ``Copula-based performance analysis for fluid antenna systems under arbitrary fading channels," {\it{IEEE Commun. Lett.}}, vol. 27, no. 11, pp. 3068-3072, Nov. 2023.

\bibitem{ref30}
Q. Zhang \emph{et al.}, “An efficient sum-rate maximization algorithm for fluid antenna-assisted ISAC system,” {\it{IEEE Commun. Lett.}}, vol. 29, no. 1, pp. 200-204, Jan. 2025.

\bibitem{ref31}
H. Niu \emph{et al.}, ``A survey on artificial noise for physical layer security: Opportunities, technologies, guidelines, advances, and trends, '' \emph{ IEEE Commun. Surv. Tutor. }, vol. 28, pp. 341-381, Sep. 2026

\bibitem{ref32}
W. K. New, K. K. Wong, H. Xu, K. F. Tong and C.-B. Chae, ``Fluid antenna system: New insights on outage probability and diversity gain," {\it{IEEE Trans. Wireless Commun.}}, vol. 23, no. 1, pp. 128-140, Jan. 2024.

\bibitem{ref33}
T. Wu {\em et al.}, ``Scalable FAS: A new paradigm for array signal processing,'' \emph{IEEE J. Sel. Topics Signal Process.}, early access, 2026.

\bibitem{ref34}
H. Xu \emph{et al.}, ``Channel estimation for FAS-assisted multiuser mmWave systems," {\it{IEEE Commun. Lett.}}, vol. 28, no. 3, pp. 632-636, Mar. 2024

\bibitem{ref35}
W. K. New \emph{et al.}, ``Channel estimation and reconstruction in fluid antenna system: Oversampling is essential," {\it{IEEE Trans. Wireless Commun.}}, vol. 24, no. 1, pp. 309-322, Jan. 2025.

\bibitem{ref36}
Y. Zhu, Z. Ding, and X. You, ``Topological perspective of large-scale multi-cell deployment of excitable waveguide dielectrics,” {\it{IEEE Wireless Commun. Lett.}}, vol. 15, pp. 151–155, 2026.

\bibitem{ref361}
T. Han, Y. Zhu, K.-K. Wong, G. Zheng, and H. Shin, ``Cell-free fluid antenna multiple access networks,” {\it{IEEE Trans. Wireless Commun.}}, vol. 24, no. 9, pp. 7237–7251, Sep. 2025.

\bibitem{ref362}
T. Han, Y. Zhu, G. Zheng and P. -D. Arapoglou, ``Fluid Antenna Enabled Compact Ultra Massive Antenna Array for Satellite Communications," {\it{IEEE J. Sel. Areas Commun.}}, vol. 44, pp. 1077-1091, 2026.

\bibitem{ref37}
J. Yao \emph{et al.}, ``Exploring fairness for FAS-assisted communication systems: from NOMA to OMA," {\it{IEEE Trans. Wireless Commun.}}, vol. 24, no. 4, pp. 3433-3449, April 2025.

\bibitem{ref38}
C. Wang \emph{et al.}, ``Fluid antenna system liberating multiuser MIMO for ISAC via deep reinforcement learning," {\it{IEEE Trans. Wireless Commun.}}, vol. 23, no. 9, pp. 10879-10894, Sep. 2024.

\bibitem{ref39}
L. Zhou \emph{et al.}, ``Fluid antenna-assisted ISAC systems," {\it{IEEE Wireless Commun. Lett.}}, vol. 13, no. 12, pp. 3533- 3537, Dec. 2024

\bibitem{ref40}
J. Yao \emph{et al.}, ``FAS-RIS communication: Model, analysis, and optimization," {\it{IEEE Trans. Veh. Technol.}}, vol. 74, no. 6, pp. 9938-9943, Jun. 2025.


\bibitem{ref41}
X. Lai \emph{et al.}, ``Revisiting spatial block-correlation model for fluid antenna systems: from constant to variable correlations," {\it{IEEE J. Sel. Areas Commun.}}, vol. 44, pp. 1335-1351, Oct. 2025.

\bibitem{ref42}
T. Wu {\em et al.}, ``Variable block-correlation modeling and optimization for secrecy analysis in fluid antenna systems,''  \emph{IEEE Trans. Wireless Commun.}, vol. 25, pp. 15069-15085, Apr. 2026.


\bibitem{ref42MA1}
L. Zhu \emph{et al.}, ``Movable antennas for wireless communication: Opportunities and challenges,” {\it{IEEE Commun. Mag.}}, vol. 62, no. 6, pp. 114–120, Jun. 2024.

\bibitem{ref42MA2}
L. Zhu, W. Ma, and R. Zhang, ``Modeling and performance analysis for movable antenna enabled wireless communications,” {\it{IEEE Trans. Wireless Commun.}}, vol. 23, no. 6, pp. 6234–6250, Jun. 2024.

\bibitem{ref42MA3}
W. Ma, L. Zhu, and R. Zhang, ``MIMO capacity characterization for movable antenna systems,” {\it{IEEE Trans. Wireless Commun.}}, vol. 23, no. 4, pp. 3392–3407, Apr. 2024.

\bibitem{ref42MA4}
L. Zhu \emph{et al.}, ``Movable antenna enabled near-field communications: Channel modeling and performance optimization,” IEEE Trans. Commun., vol. 73, no. 9, pp. 7240–7256, Sep. 2025.


\bibitem{ref43}
C. Zhang \emph{et al.} ``Fluid antenna-aided robust secure transmission for RSMA-ISAC systems," {\it{arXiv preprint}}, 2025.

\bibitem{ref44}
F. R. Ghadi\emph{et al.} ``Fluid antenna-aided rate-splitting multiple access," {\it{IEEE Trans. Veh. Technol.}}, vol. 75, no. 2, pp. 3417–3422, Aug. 2025.

\bibitem{ref45}
F. R. Ghadi \emph{et al.} ``Phase-mismatched STAR-RIS with FAS-assisted RSMA users," \emph{IEEE Trans. Commun.}, vol. 74, pp. 2032–2046, Dec. 2025.

\bibitem{ref46}
F. R. Ghadi \emph{et al.} ``UAV-relay assisted RSMA fluid antenna system: Outage probability analysis," \emph{IEEE Wireless Commun. Lett.}, vol. 14, no. 9, pp. 2907–2911, Sep. 2025.

\bibitem{ref461}
J. Liu \emph{et al.},  ``FAS-RSMA: Can fluid antennas elevate RSMA performance?," \emph{IEEE Trans. Commun.}, vol. 74, pp. 6476-6492, Mar. 2026.

\bibitem{ref47}
J. Liu \emph{et al.}, ``Fluid Antennas Meet Rate-Splitting Multiple Access: A New Path Forward for 6G Networks," {\it{arXiv preprint}}, Apr. 2026.

\bibitem{ref48}
I. S. Gradshteyn and I. M. Ryzhik, {\it{Table of Integrals, Series, and Products}}, 7th ed. San Diego, CA: Academic, 2007.
\end{thebibliography}
\end{document}